\documentclass[referee]{raa}           

\usepackage{graphicx,times}
\usepackage{natbib}
\usepackage{amssymb,amsmath}
\bibpunct{(}{)}{;}{a}{}{,}

\usepackage[pagebackref=true]{hyperref}

\newcommand{\obj}{IL~Leo}

\begin{document}

   \title{The polar IL~Leo in a low accretion state}
%   \subtitle{I. Place Your Subtitle Here}

   \volnopage{Vol.0 (20xx) No.0, 000--000}      %%preserved for Editor. DOn't remove!
   \setcounter{page}{1}          %%starting page, preserved for Editor. DOn't remove!

   \author{M. V. Suslikov        %% Put your Chinese name in "( )" if you like. Note to open line 11 "\usepackage[UTF8]{ctex}"
      \inst{1,2}
   \and A. I. Kolbin
      \inst{2}
   \and N. V. Borisov
      \inst{2}
   }

   \institute{Kazan Federal University, Kazan 420008, Russia; {\it mvsuslikov@outlook.com}\\
        \and
             Special Astrophysical Observatory, Russian Academy of Sciences, Nizhnii Arkhyz 369167, Russia\\
\vs\no
   {\small Received 20xx month day; accepted 20xx month day}}

\abstract{ We performed an optical study of the magnetic period-bouncer candidate {\obj}. Long-term photometric analysis over $\approx 20$ years reveals multiple state transitions. Modelling the ultraviolet and optical spectral energy distribution refined the white dwarf parameters, yielding a mass of $M_\textrm{wd} = 0.74 \pm 0.05~M_{\odot}$ and an effective temperature of $T_\mathrm{eff} = 12700 \pm 360$~K. We analyzed phase-resolved spectroscopy obtained with the 6-m BTA telescope and the VLT during the low state. Orbital variability of the H$\alpha$ emission, inferred from dynamical spectra and Doppler tomograms, suggests that it originates in the accretion stream. Zeeman splitting gives a mean magnetic field of $B = 40.7 \pm 0.5$~MG. Modelling two sets of cyclotron spectra determined a low-state accretion rate of $\dot{M} = (2.5 - 4.1) \times 10^{-13}~M_{\odot}$~yr$^{-1}$ and a magnetic field of $B_\mathrm{m} \approx 41$~MG near magnetic pole.
}

   \authorrunning{M. V. Suslikov, A. I. Kolbin \& N. V. Borisov}            %author_head in even pages
   \titlerunning{The polar IL~Leo in a low accretion state}                 % title_head in odd pages
   
   \maketitle

%________________________________________________ sections below
% 
\section{Introduction}           %% first-level sections will be auto-capitalized
\label{sec:intro}

% Authors can give a citation as `\citealt{Michel+etal+1992}'.
% You may also use \cite, \citep and \citet for citation, and use Table~1
% or Figure~1 and so forth. Using \ref and \label for cross-references of
% Tables/Figures is a good way in adjusting/adding/removing text, tables or
% figures.

Cataclysmic variables are close binary systems consisting of an accreting white dwarf and a low-mass donor star that fills its Roche lobe \citep{Warner1995, Hellier2001}. A distinct subclass of these binaries is formed by polars, or AM Her-type stars, which are characterized by strongly magnetized white dwarfs with field strengths of $B \sim 10-100$~MG. In contrast to non-magnetic cataclysmic variables, polars do not possess accretion disks. The accretion stream leaving the L$_1$ Lagrangian point encounters the threading region, where its ram pressure becomes comparable to the magnetic pressure. Beyond this region, the ionized gas is channeled along the magnetic field lines toward the magnetic poles of the accretor \citep{Cropper1990}. The impact of the infalling material on the white dwarf surface forms hot accretion spots ($T \sim 10$~keV), which are sources of hard X-ray bremsstrahlung and cyclotron emission observed in the optical and infrared ranges. Due to the strong magnetic coupling, polars are synchronized systems in which the spin period of the white dwarf is equal to the orbital period.

Polars are known to switch between different accretion states, manifested as irregular variations in the system's mean brightness by several magnitudes on timescales ranging from a few days to years \citep{Duffy2022}. In the high state, the accretion rate reaches $\dot{M} \sim 10^{-11} - 10^{-9}~M_{\odot}$~yr$^{-1}$, leading to the formation of a standing shock above the white dwarf surface, with the post-shock cooling primarily through X-ray emission. In the low state, the accretion rate decreases to  $\dot{M} \sim 10^{-13} - 10^{-12}~M_{\odot}$~yr$^{-1}$ \citep{Hessman2000}, which is insufficient to sustain a shock. In this regime, cyclotron cooling dominates the energy losses from the accretion region \citep{Beuermann1994}. A distinctive feature of low states is the presence of bright cyclotron harmonics in the optical and infrared spectra \citep{Campbell2008_3}. The physical mechanism driving these state transitions remains debated but is generally attributed to changes in the donor star's magnetic field configuration, which can partially or completely suppress mass transfer \citep{Livio94, King1998}.

A related but relatively small group of systems, known as LARPs (Low Accretion Rate Polars; \citealt{Schwope2002}), exhibits observational properties similar to those of polars in low states. However, LARPs do not undergo state transitions and, unlike most polars, have long orbital periods of several hours. These systems have been shown to be detached, with accretion onto the white dwarf proceeding via the stellar wind of the M-dwarf secondary \citep{Webbink2005}. Polars are thought to evolve from LARPs through angular momentum loss, which is why these objects are also referred to as pre-polars (\citealt{Schwope2009}).

A notable finding is the identification of LARP-like objects among short-period systems with $P_\mathrm{orb} \approx 80$~min, corresponding to the minimum orbital period for cataclysmic variables. These objects are considered to be period-bouncers, representing a late evolutionary stage in which the donor has evolved into a brown dwarf. The most well-studied examples of such systems are EF~Eri \citep{Beuermann1987, Schwope2007, Khangale2025} and V379~Vir \citep{Schmidt2005, Debes2006, Farihi2008, Stelzer2017, Suslikov2025a, Suslikov2025b}. Their donors are thought to lose mass primarily through the L$_1$ Lagrangian point, similar to classical polars, while the low accretion rates are explained by the slow shrinkage of the Roche lobe expected for such old systems \citep{Knigge2011}. Investigating these systems is crucial for understanding the late-stage evolution of magnetic cataclysmic variables and the nature of their magnetic fields. Recent work by \cite{Schreiber2023} highlights the key role of magnetic fields in interpreting the observed population of period-bouncers.

The variable {\obj} (also known as SDSS~J103100.55+202832.1) was first classified by \citet{Schmidt2007} as a polar with a low accretion rate. Its spectra exhibit cyclotron harmonics corresponding to a magnetic field of $B = 42$~MG, and a short orbital period of $P_\mathrm{orb} \approx 82$~min was measured. Based on the low estimated accretion rate of $\dot{M} \sim 10^{-13}~M_{\odot}$~yr$^{-1}$, the authors suggested that mass transfer occurs via the stellar wind of the donor rather than through Roche-lobe overflow. Ultraviolet observations with GALEX \citep{Linnell2010} revealed photometric variability that can be modeled with a hot spot of $T_\mathrm{eff} \approx 13000$~K on the white dwarf surface. Phase-resolved optical spectroscopy \citep{Parsons2021} showed variability of the cyclotron harmonics, weak $H\alpha$ emission, and Zeeman absorption features. From X-ray observations with the eROSITA telescope onboard the Spektr--RG observatory, \citet{Giraldo2024} determined a flux of $F_\mathrm{x} = (6.71 \pm 0.55) \times 10^{-13}$~erg~s$^{-1}$~cm$^{-2}$. A recent analysis of ZTF photometry \citep{vanRoestel2025} demonstrated that the system can switch to a higher accretion state. This indicates that {\obj} is a semi-detached binary system. In its overall observational characteristics, {\obj} closely resembles the polar EF~Eri, which has remained in a low state for an extended period.
In this study, we present a detailed investigation of {\obj} based on multiwavelength photometric data combined with phase-resolved spectroscopy acquired with the VLT and the 6-m BTA telescope of the Special Astrophysical Observatory of the Russian Academy of Sciences.

\section{Observations and data reduction}
\label{sec:obs}

\subsection{BTA spectroscopy}
\label{sec:bta_spec}

Spectroscopic observations of {\obj} were obtained during the night of 26/27 February 2023 with the 6-m BTA telescope at the Special Astrophysical Observatory of the Russian Academy of Sciences (SAO RAS). The telescope was equipped with the SCORPIO--1 focal reducer \citep{Afanasiev2005}, operated in the long-slit spectroscopy mode. The configuration using the VPHG550G volume-phase holographic grating (550~lines\,mm$^{-1}$) and a slit of width $1.2''$ provided continuous wavelength coverage from 3800 to 7500~\AA\ at a spectral resolution of $\Delta\lambda \approx 12$~\AA. A total of 18 spectra with individual exposure times of 300~s were acquired, spanning the full orbital period of {\obj}. The phase-resolved spectroscopic data were reduced using standard long-slit procedures within the \textsc{PyRAF} package\footnote{The PyRAF package for reduction and analysis of astronomical data is available at \url{https://iraf-community.github.io/pyraf.html}}. The preliminary processing included bias subtraction, removal of cosmic rays, and flat-field correction. Wavelength calibration and correction for geometric distortions were performed using spectra of a He--Ne--Ar arc lamp. One-dimensional spectra were extracted using the optimal extraction algorithm \citep{Horne1986}, with subtraction of the sky background. Flux calibration was carried out with the spectrophotometric standard star HZ~44, which was additionally used to remove telluric absorption features. The times of the individual exposures were converted to mid-exposure Barycentric Julian Dates (BJD), and barycentric corrections were applied to the radial velocities.

\subsection{Archival VLT spectra}
\label{sec:vlt_spec}

To ensure a comprehensive analysis, we incorporated archival data retrieved from the ESO Science Archive\footnote{The ESO Science Archive Facility is available at \url{https://archive.eso.org/}}. Phase-resolved spectroscopy of {\obj} was performed on 16/17 April 2017 with the 8-m VLT/UT2 telescope at the Paranal Observatory, using the medium-resolution echelle spectrograph X-shooter. The instrument splits the incoming light into three spectral ranges: UVB (300--560 nm), VIS (560--1024 nm), and NIR (1024--2480 nm) --- each recorded independently by the corresponding spectrograph arm. These observations were originally published by \citet{Parsons2021} and are summarized in Table~\ref{table:obs_log}. We reduced the raw data using the X-shooter pipeline within the \textsc{EsoReflex} package (version 2.11.5). VIS spectra were corrected for telluric absorption using \textsc{Molecfit} \citep{Smette2015, Kausch2015}. The UVB and VIS spectra were obtained with different exposure times, resulting in slightly different observation phases. To merge the spectra at matching phases, cubic-spline interpolation grids were constructed. The NIR spectra were found to be relatively noisy and were therefore excluded from further analysis. The log of spectroscopic observations is presented in Table~\ref{table:obs_log}.

\begin{table*}
   \caption{Log of spectroscopic observations of {\obj}. The table lists the telescopes and instruments, observation dates, total duration, number of spectra acquired ($N$), spectral coverage, and exposure times ($\Delta t_{\mathrm{exp}}$).}
   \label{table:obs_log}
   \vspace*{-3mm}
   \begin{center}
       \begin{tabular}{lccccc}
           \hline
           Telescope /         & Date               & Duration  & $N$ & Spectral range & $\Delta t_{\mathrm{exp}}$, \\
           Instrument          & (UT)               & (min)     &     & (\AA)          & (s)                        \\ \hline
           BTA/SCORPIO-1       & 26/27 Feb. 2023    & 91        & 18  & 3800--7300     & 300                        \\
           VLT/X-shooter UVB   & 16/17 Apr. 2017    & 115       & 20  & 3000--5600     & 300                        \\
           VLT/X-shooter VIS   & 16/17 Apr. 2017    & 115       & 18  & 5600--10000    & 334                        \\
           VLT/X-shooter NIR   & 16/17 Apr. 2017    & 114       & 22  & 10000--24000   & 300                        \\
           \hline
       \end{tabular}
   \end{center}
\end{table*}

\subsection{Photometric data}
\label{sec:phot_data}

To investigate the long-term optical variability of {\obj}, we utilized data from photometric sky surveys including CSS (Catalina Sky Survey; \citealt{Drake2009}), ZTF (Zwicky Transient Facility; \citealt{Masci2019}), Pan-STARRS (Panoramic Survey Telescope and Rapid Response System; \citealt{Flewelling2020}), PTF (Palomar Transient Factory; \citealt{Low2009}), and Gaia DR3 \citep{Vallenari2023}.

As reported by \cite{Linnell2010}, the system was observed with the \textit{GALEX} ultraviolet telescope in the $FUV$ and $NUV$ bands in February 2008, with a total exposure time of 11136~s. Data extraction from the photon event archive and aperture photometry were performed using the \textsc{gPhoton} software package \citep{Million2016}. The resulting light curves were binned with a time interval of 300~s. Additionally, ultraviolet fluxes were obtained from archival \textit{Swift} UVOT observations in the $UVM2$ and $UVW1$ filters, with aperture photometry carried out using the \texttt{uvotmaghist} task within the \textsc{HEASoft} package\footnote{The HEASoft software package is available at \url{https://heasarc.gsfc.nasa.gov/lheasoft/}}.

Despite the relative proximity of {\obj}, we estimated the line-of-sight extinction using three-dimensional interstellar dust maps \citep{Green2018} accessed via the \textsc{dustmaps} Python package\footnote{The \textsc{dustmaps} package is available at \url{https://dustmaps.readthedocs.io/}}. Adopting the system coordinates and a distance of $d = 518 \pm 61$~pc, derived from the Gaia DR3 parallax \citep{Vallenari2023}, the color excess was found to be $E(B-V) = 0.0238^{+0.0002}_{-0.0004}$. The observed photometric and spectroscopic fluxes, subsequently used in modelling, were corrected for interstellar extinction following \cite{Fitzpatrick1999}.

\section{Photometric analysis}
\label{sec:phot_analys}

\subsection{Long-term variability}
\label{sec:longterm}

Using photometric survey data, we constructed an optical light curve covering approximately 20 years of observations of {\obj}. Beyond the orbital variability, the source exhibits long-term brightness changes of $\sim 2 - 3^\mathrm{m}$ on timescales of several years. We distinguish a low state with an average magnitude of $\langle g \rangle \sim 20^\mathrm{m}$, a high state ($\langle g \rangle \sim 17^\mathrm{m}$), and an intermediate state ($\langle g \rangle \sim 18.5^\mathrm{m}$). Such an amplitude of variability is typical for polars and reflects changes in the mass-accretion rate within the system.

\begin{figure*}[h!]
   \centering 
   \includegraphics[width=\textwidth]{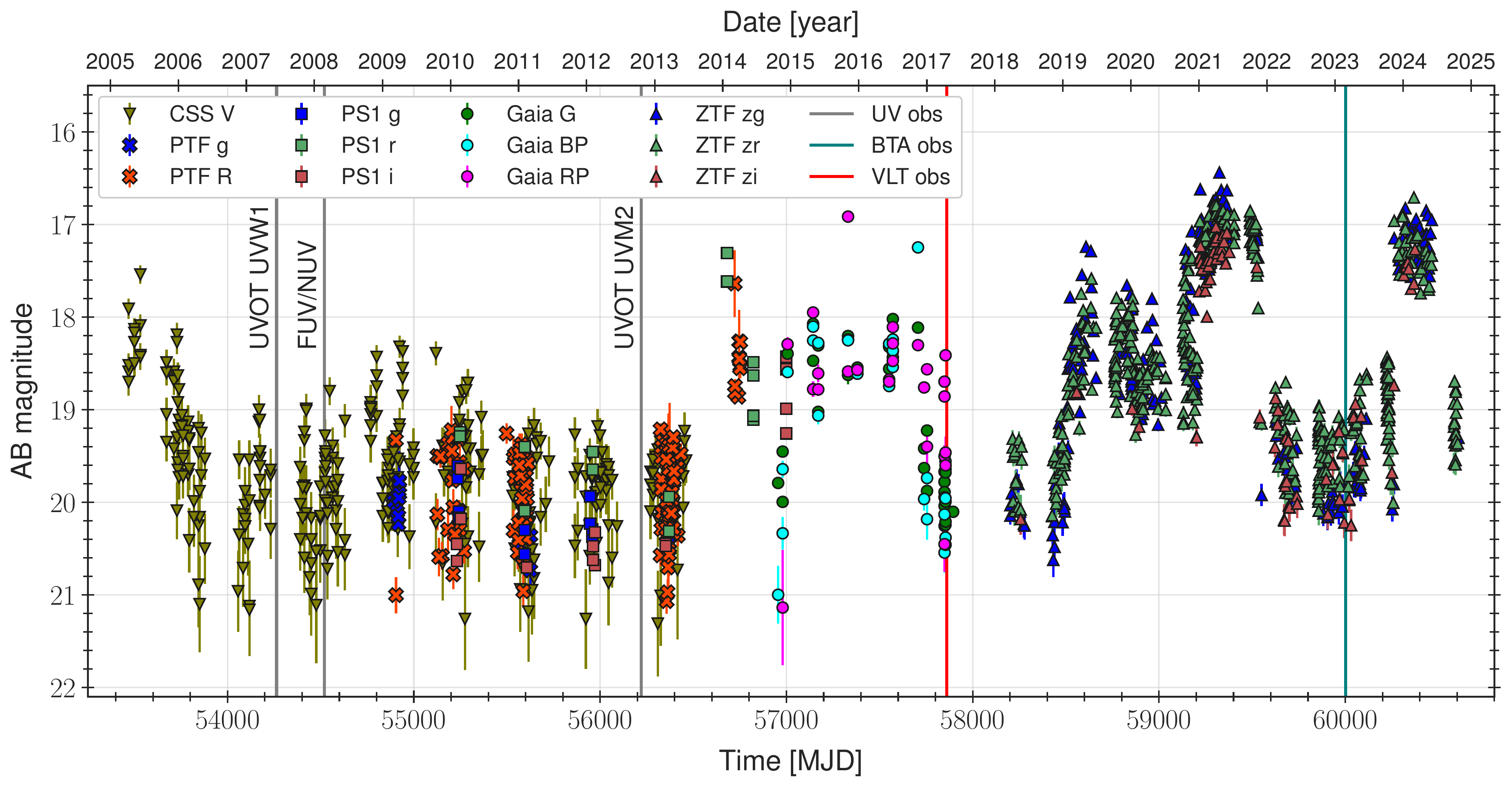} 
   \caption{Long-term optical light curve of {\obj}. Photometric data are presented from ZTF (bands $g$, $r$, $i$), PTF (bands $g$, $R$), CSS (band $V$), Gaia (bands $G$, $BP$, $RP$), and Pan-STARRS (bands $g$, $r$, $i$) catalogs. Vertical lines indicate the epochs of the \textit{GALEX} and \textit{Swift} UVOT ultraviolet observations and the BTA and VLT spectroscopic observations.}
   \label{fig:longterm_lc}
\end{figure*}

\subsection{Rotational variability}
\label{sec:rot_var}

\begin{figure}[h!]
   \centering
   \includegraphics[width=1.0\linewidth]{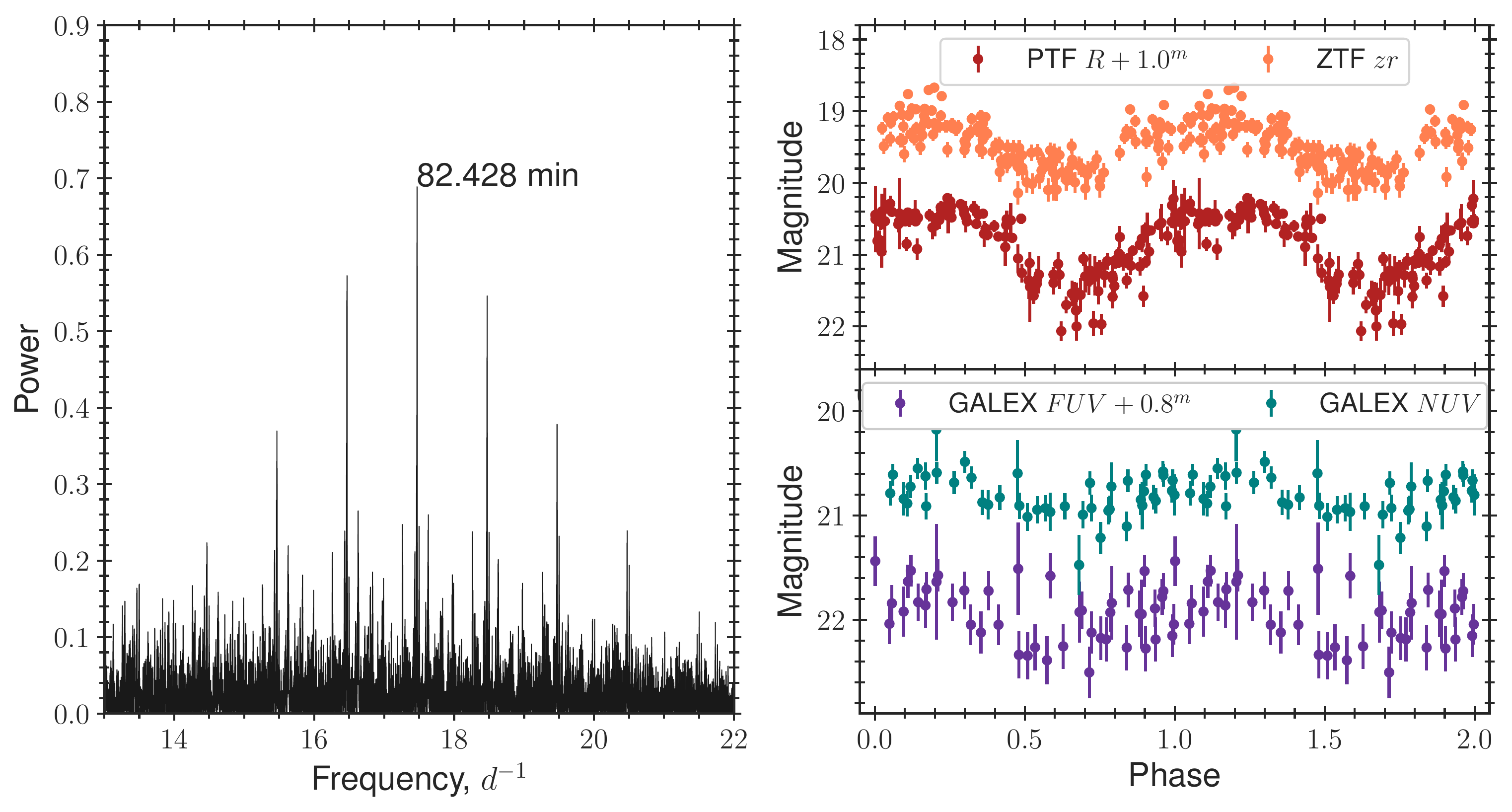}
   \caption{
   Left panel: Lomb--Scargle periodogram derived from the low-state photometry of {\obj} in the ZTF $zr$ and PTF $R$ bands. The highest power peak at a frequency of $17.4699 \pm 0.0004$~d$^{-1}$ corresponds to a period of $82.428 \pm 0.001$~min. Right panel: phase-folded light curves of {\obj} in the low state for the optical ZTF $zr$ and PTF $R$ bands and the ultraviolet GALEX $FUV$ and $NUV$ bands.
   }
   \label{fig:phase_folded}
\end{figure}

To refine the orbital period, we selected photometric data of {\obj} corresponding to its low state. The strongest optical variability appears in the ZTF $zr$ and PTF $R$ bands, which are dominated by rotational modulation of the 4th cyclotron harmonic. To compute the periodogram based on the two time series, we applied the multiband Lomb--Scargle method. The most significant peak in the power spectrum (Fig.~\ref{fig:phase_folded}) corresponds to a period of $82.428 \pm 0.001$ min, with the uncertainty estimated from the half-width of the peak. The refined ephemeris is given by:
\begin{gather}\label{eq:ephemeris}
    \mathrm{BJD} = 2454904.7205(4) + 0.0572413(7) \times E,
\end{gather}
where the reference epoch is defined as the presumed inferior conjunction of the donor star and was derived from modelling the H$\alpha$ line dynamic spectrum (see Section \ref{sec:ha_emis}).

Figure~\ref{fig:phase_folded} also presents the phase-folded light curves in the GALEX ultraviolet bands $FUV$ and $NUV$. Comparison with the optical light curves reveals that the brightness variations caused by the rotation of the hot spot on the white dwarf coincide with the variability of the cyclotron source.

\subsection{Spectral energy distribution}
\label{sec:sed}

The parameters of the white dwarf in {\obj} were inferred from its spectral energy distribution (SED) in the low state. At the minima of the system's brightness, when the accretion spot is partially eclipsed, the ultraviolet and optical emission is dominated by the white dwarf. To model the SED, we used the minimum observed fluxes: GALEX ($FUV$, $NUV$) data from the phase-folded light curves (averaged over $\phi = 0.5 - 0.7$; see Fig.~2), \textit{Swift} UVOT ($UVM2$, $UVW1$) fluxes, as well as ZTF ($z$, $r$) and Pan-STARRS ($g$) data, which are not affected by strong cyclotron harmonics.

\begin{figure}[h!]
    \centering
    \includegraphics[width=0.8\linewidth]{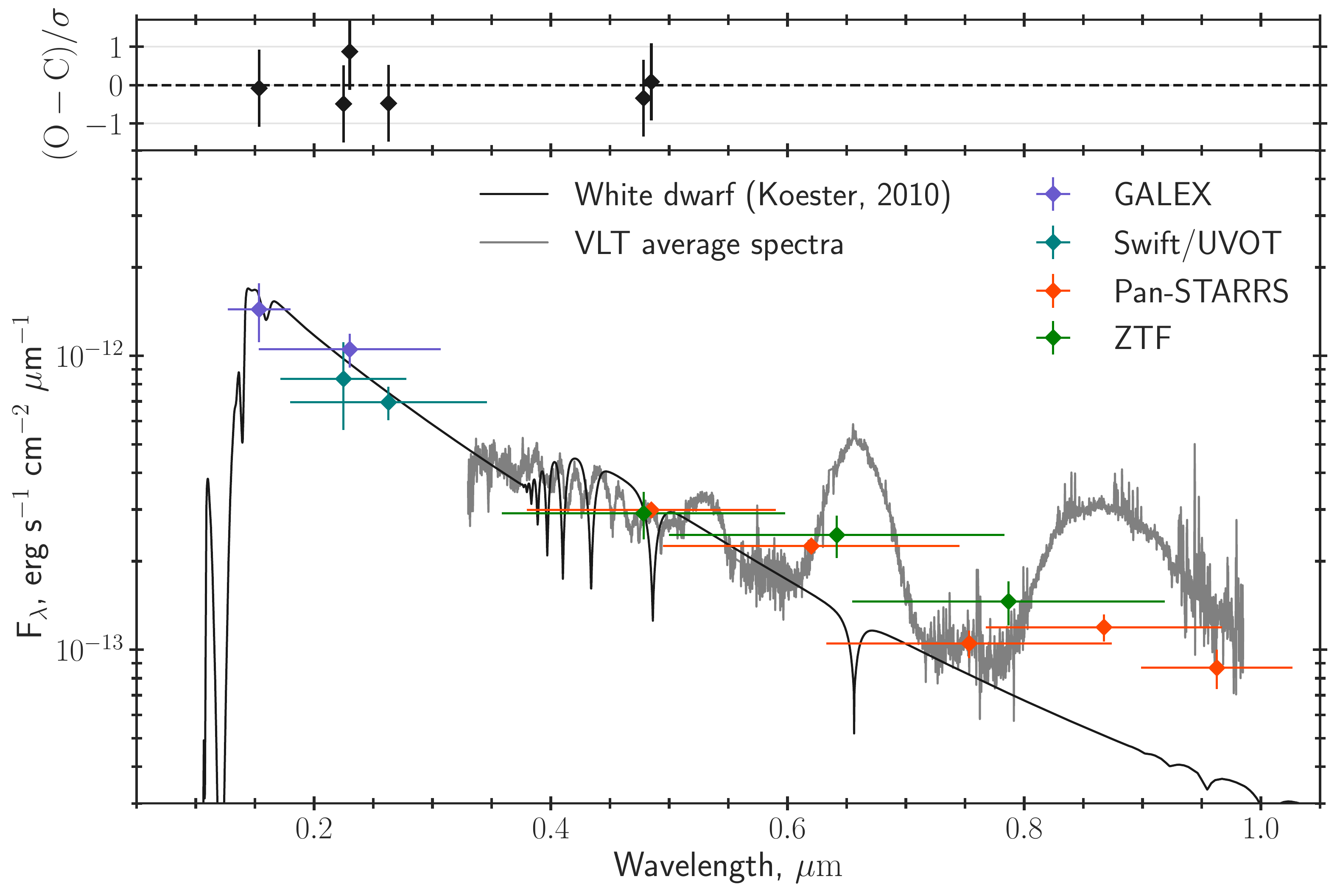}
    \caption{Bottom panel: the spectral energy distribution of {\obj} in the low state. The plot presents the observed photometric fluxes (with horizontal bars indicating the effective bandwidths), the averaged VLT spectrum, and the white dwarf model spectrum. Top panel: the $(O - C)/\sigma$ diagram showing the residuals between the observed fluxes and the synthetic fluxes used in the spectral fitting.}
    \label{fig:sed}
\end{figure}

The observed fluxes were modeled using plane-parallel LTE hydrogen white dwarf atmospheres \citep{Koester2010}. Grids of theoretical spectra were computed over a wide range of effective temperatures $T_\mathrm{eff}$ and surface gravities $\log g$. The model spectrum was then fitted by minimizing the
\begin{equation}\label{eq:sed_fit}
    \chi^2 = \displaystyle\sum_{k} \Big(\frac{f_{k}^{\rm obs} - \frac{1}{4} \theta_\mathrm{wd}^2 f_{k}^{\rm wd}}{\sigma_k}\Big)^2,
\end{equation}
where $f_{k}^{\rm obs}$ is the extinction-corrected observed flux in the $k$-th photometric band, $f_{k}^{\rm wd}$ is the corresponding theoretical flux at the stellar surface, $\theta_\mathrm{wd}$ is the angular diameter of the white dwarf, and $\sigma_k$ is the uncertainty of the observed flux. The model fluxes $f_{k}^{\rm wd}$ were calculated by convolving the interpolated model spectrum with the transmission curve of the respective photometric filter. The angular diameter $\theta_\mathrm{wd}$ was determined by minimizing equation \ref{eq:sed_fit}. Parameter optimization was performed using the Nelder--Mead method \citep{Gao2012}, and uncertainties were estimated via Monte Carlo simulations. The spectral energy distribution of {\obj} is best reproduced by a white dwarf atmosphere with an effective temperature of $T_\mathrm{eff} = 12700 \pm 360$~K and a surface gravity of $\log g = 8.2 \pm 0.4$ (Fig.~\ref{fig:sed}). Combining the derived angular diameter $\theta_\mathrm{wd}$ with the Gaia DR3 parallax, we determine the radius of the accretor to be $R_\mathrm{wd} = 0.0107 \pm 0.0006~R_\odot$. Applying the white dwarf mass--radius relation \citep{Nauenberg72} yields a mass of $M_\mathrm{wd} = 0.74 \pm 0.05~M_\odot$, corresponding to $\log g = 8.25$, in agreement within uncertainties with the value derived from the SED fit. The flux excess at wavelengths $\lambda > 0.8~\mu$m likely arises from a combination of cyclotron emission from the accretion spot and contribution from the secondary component. The analysis of the phase-averaged VLT infrared spectrum, despite substantial telluric contamination and noise, suggests that the donor flux is unlikely to exceed $0.5 \times 10^{-17}$~erg~s$^{-1}$~cm$^{-2}$ within the $K$ band. Based on the theoretical relation \citep{Knigge2011}, for a system with an orbital period of $P_\mathrm{orb} \sim 82$~min, the donor's spectral type should be later than M5.5. However, due to the presence of cyclotron emission and the weak infrared flux, the available data do not allow robust conclusions regarding the donor's degeneracy.

\section{Phase-resolved spectroscopy analysis}
\label{sec:spec_analys}

\subsection{Dynamic spectra}
\label{sec:dyn_spec}

According to Fig.~\ref{fig:longterm_lc}, the optical spectroscopic observations of {\obj} at the BTA and VLT telescopes were conducted when the system was in a low state. In the optical range, the emission is dominated by the white dwarf and the accretion spot. The most prominent spectral features are the broad cyclotron harmonics with maxima near 8600~\AA\ (3rd harmonic), 6600~\AA\ (4th harmonic), and 5300~\AA\ (5th harmonic). The VLT NIR spectra also reveal the 2nd cyclotron harmonic at a wavelength of $\lambda \approx 12200$~\AA. Both the position and intensity of the cyclotron harmonics are modulated by the rotation of the white dwarf. In addition, the blue part of the spectrum shows Zeeman-split absorption components of the $H\beta$ line, formed in the white dwarf photosphere. A comparison of the phase-resolved spectroscopy from the two data sets indicates that the BTA observations correspond to a higher accretion rate, as evidenced by the stronger H$\alpha$, H$\beta$, H$\gamma$, and HeI(5876) emission lines and by the higher continuum flux. The cyclotron harmonics in the BTA observations appear shifted toward longer wavelengths and are broader than those in the VLT data. This apparently reflects either an increase in the local accretion rate or a change in the magnetic field strength in the cyclotron-emitting region, possibly caused by a variation in its location on the white dwarf's surface.

\begin{figure*}[h!]
    \centering 
    \includegraphics[width=\textwidth]{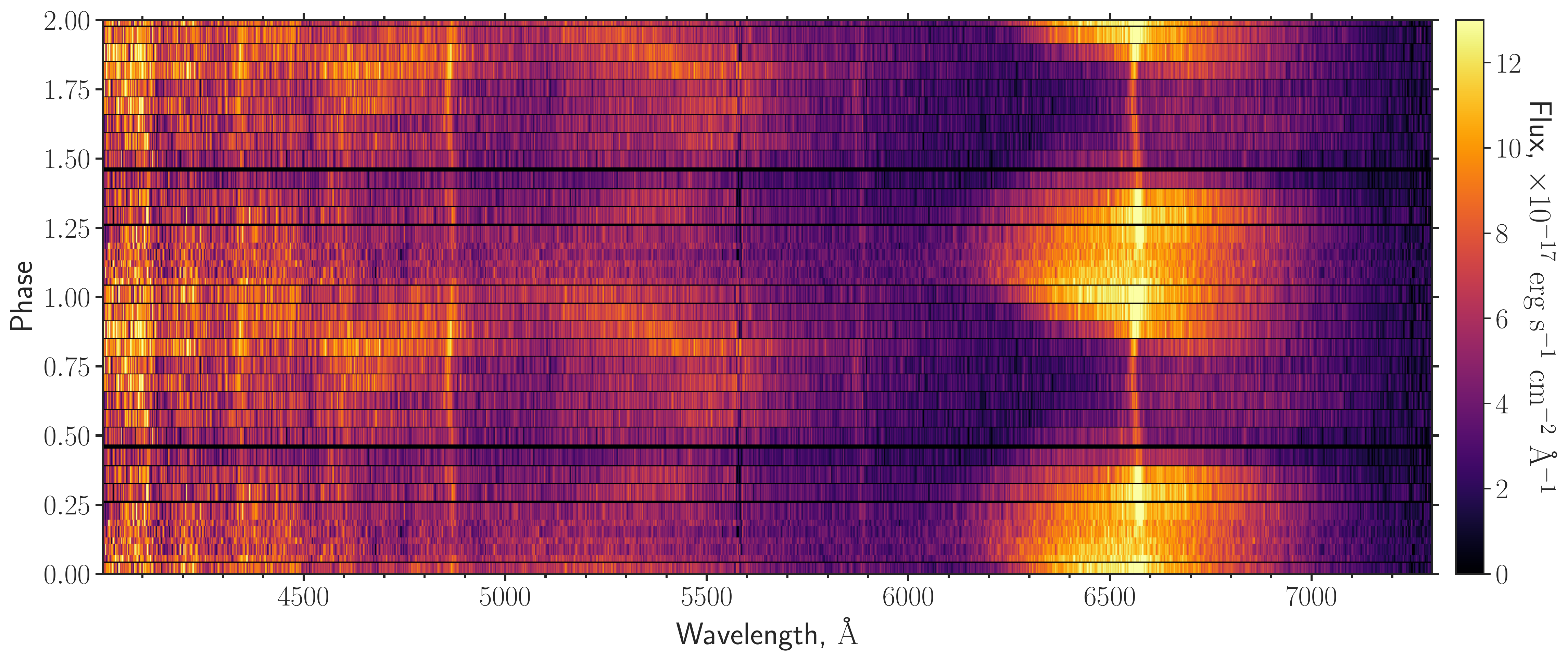} 
    \caption{Optical dynamic spectrum of {\obj} constructed from BTA observations. The 4th and 5th cyclotron harmonics appear at 6600~\AA\ and 5400~\AA, respectively. In the blue part of the spectrum ($\lambda < 5000$~\AA), the Zeeman components of the photospheric H$\beta$ line can be seen.}
    \label{fig:dynspec_bta}
\end{figure*}

\subsection{H$\alpha$ emission}
\label{sec:ha_emis}

An inspection of the two spectral datasets shows that the Balmer line intensities in the BTA observations are significantly higher than in the VLT data, where only weak $H_{\alpha}$ emission is detected. From the phase-averaged BTA and VLT spectra, the $H_{\alpha}$ line flux ratio is $F_{H\alpha}^{BTA} / F_{H\alpha}^{VLT} \approx 9.3$. The BTA spectra also exhibit clear H$\beta$ and H$\gamma$ emission lines, while these features are nearly absent in the VLT data. Based on their analysis of the VLT spectra, \cite{Parsons2021} derived a radial velocity semi-amplitude of $K = 384 \pm 5$ km~s$^{-1}$ for the H$\alpha$ line.

\begin{figure}[h!]
   \centering 
   \includegraphics[width=\linewidth]{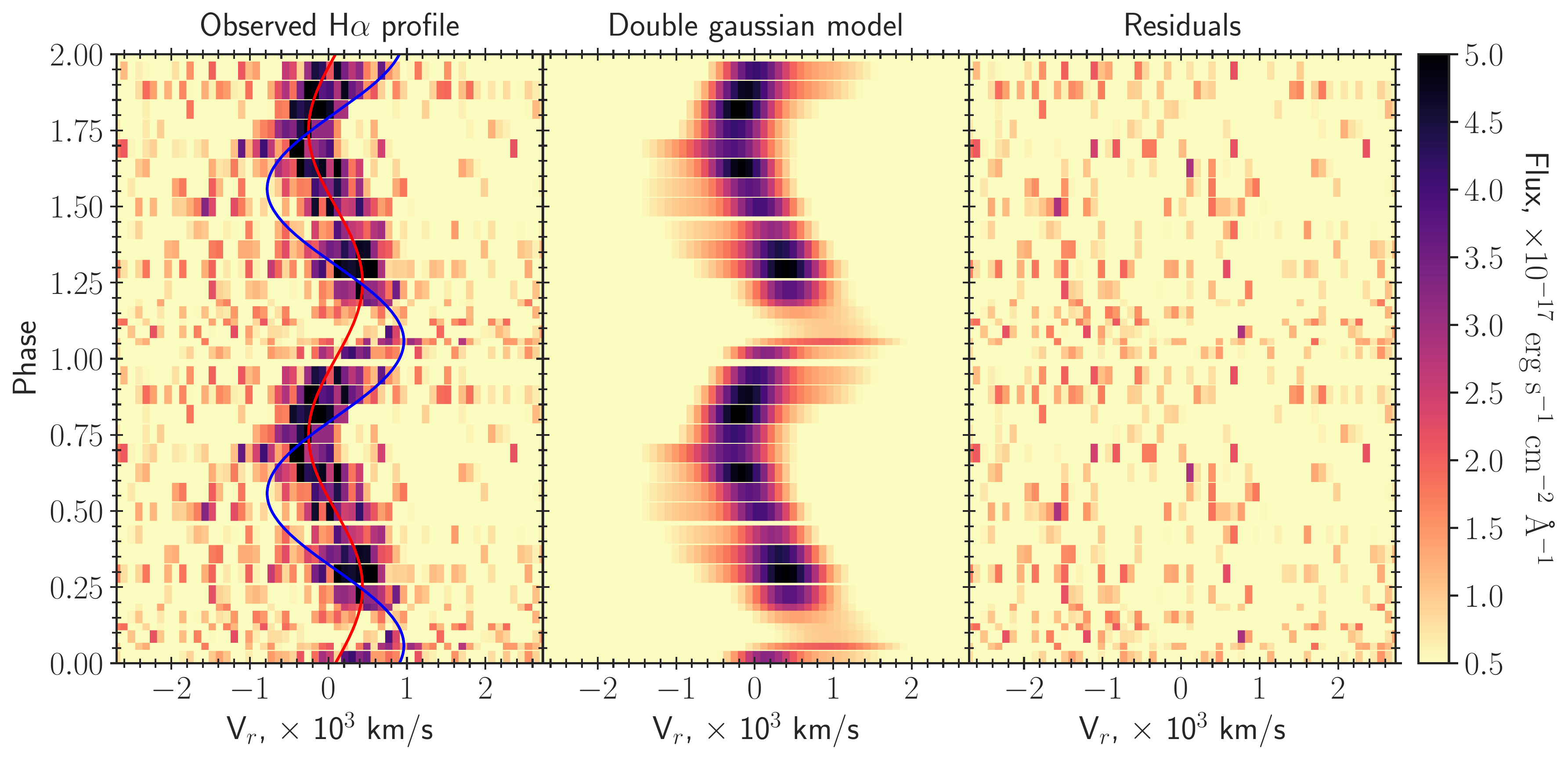}
   % \vspace*{-2mm}
   \caption{Left panel: the observed dynamic spectrum of the H$\alpha$ line with superimposed sinusoids tracing the centroids of the bright narrow (red) and faint broad (blue) components. Middle panel: the composite emission source model fitted with two Gaussians. Right panel: the residuals between the observed and model dynamic spectra.}
   \label{fig:dynspec_ha}
\end{figure}

To identify the regions that produce the emission lines, we analyzed the orbital variability of the H$\alpha$ emission profile in the BTA data. The H$\alpha$ dynamical spectrum was modeled as the sum of two Gaussian components whose radial velocities vary with orbital phase according to a sinusoidal law with the white dwarf's spin period (see Fig.~\ref{fig:dynspec_ha}). The Gaussian widths were assumed to be constant, while their amplitudes were fitted independently for each individual spectrum. Although the emission behavior in polars can be more complex (see, e.g., \citealt{Schwope1997}), a two-component model with phase-modulated Gaussians of fixed width is often sufficient for low-resolution data \citep{Kolbin2023, Liu2023, Lin2025}. The parameters of the components were determined using a least-squares fitting procedure. The resulting model reproduces the observed H$\alpha$ dynamical spectrum as a combination of a bright narrow component with a radial velocity semi-amplitude of $K_n = 344 \pm 21$ km~s$^{-1}$ and a width of $\sigma_n = 300 \pm 13$ km~s$^{-1}$, and a faint broad component with $K_b = 871 \pm 63$ km~s$^{-1}$ and $\sigma_b = 589 \pm 38$ km~s$^{-1}$, for a systemic velocity of $\gamma = 91 \pm 12$ km~s$^{-1}$\footnote{It should be noted that radial velocities may be uncertain by several tens of km~s$^{-1}$ due to uneven illumination of the SCORPIO-1 slit.}. The reference phase was defined under the assumption that the narrow component originates on the donor star near the inner Lagrange point $L_1$. Figure~\ref{fig:dynspec_ha} shows the corresponding sinusoids describing the phase modulation of the Gaussian centroids, which exhibit a phase shift of $\Delta\phi \approx 0.19 \pm 0.01$. This two component model provides a satisfactory description of the observed H$\alpha$ dynamical spectrum, yielding a reduced chi-squared value of $\chi^2_\nu = 1.2$. For comparison, a single sinusoidally varying Gaussian component results in a slightly poorer fit, with $\chi^2_\nu = 1.3$.

Some insight into the mechanism responsible for the formation of the emission lines can be obtained from the Balmer decrement. In addition to the H$\alpha$ emission, the BTA spectra also show the H$\beta$ and H$\gamma$ lines. From the phase-averaged spectrum, we measured the total fluxes in the Balmer lines: $F_{H\alpha} = 7.61 \times 10^{-16}$~erg~s$^{-1}$~cm$^{-2}$, $F_{H\beta} = 4.63 \times 10^{-16}$~erg~s$^{-1}$~cm$^{-2}$, and $F_{H\gamma} = 3.3 \times 10^{-16}$~erg~s$^{-1}$~cm$^{-2}$. The resulting flux ratios $F_{H\alpha} / F_{H\beta} = 1.64$ and $F_{H\gamma} / F_{H\beta} = 0.72$ may indicate that most of the Balmer emission is produced in a partially thermalized region and in a moderately optically thick region.
 
\subsection{Doppler tomography}
\label{sec:dop_tomog}

\begin{figure}[h]
   \begin{minipage}[t]{0.495\linewidth}
   \centering
    \includegraphics[width=0.8\linewidth]{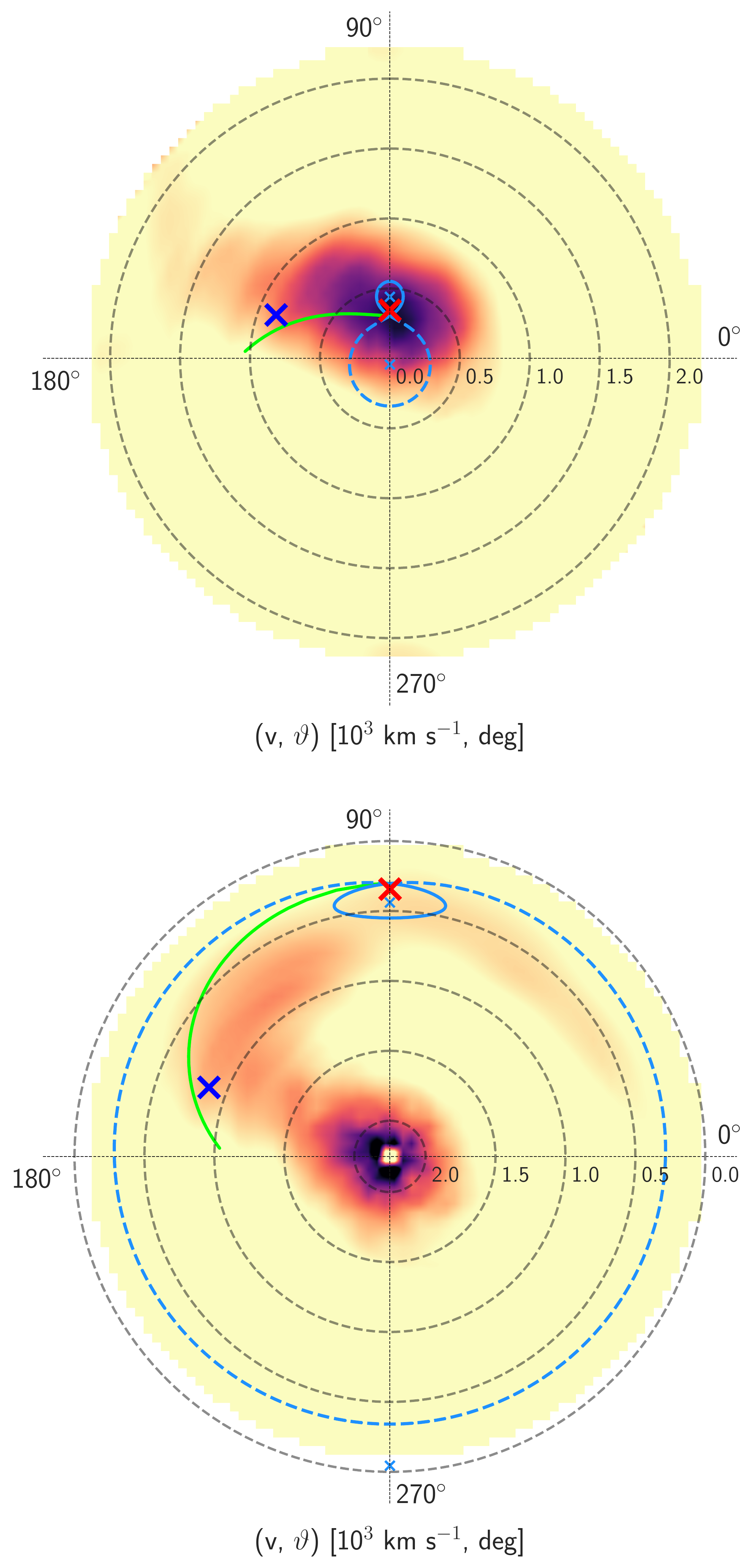}
   \end{minipage}%
   \begin{minipage}[t]{0.495\linewidth}
   \centering
    \includegraphics[width=0.8\linewidth]{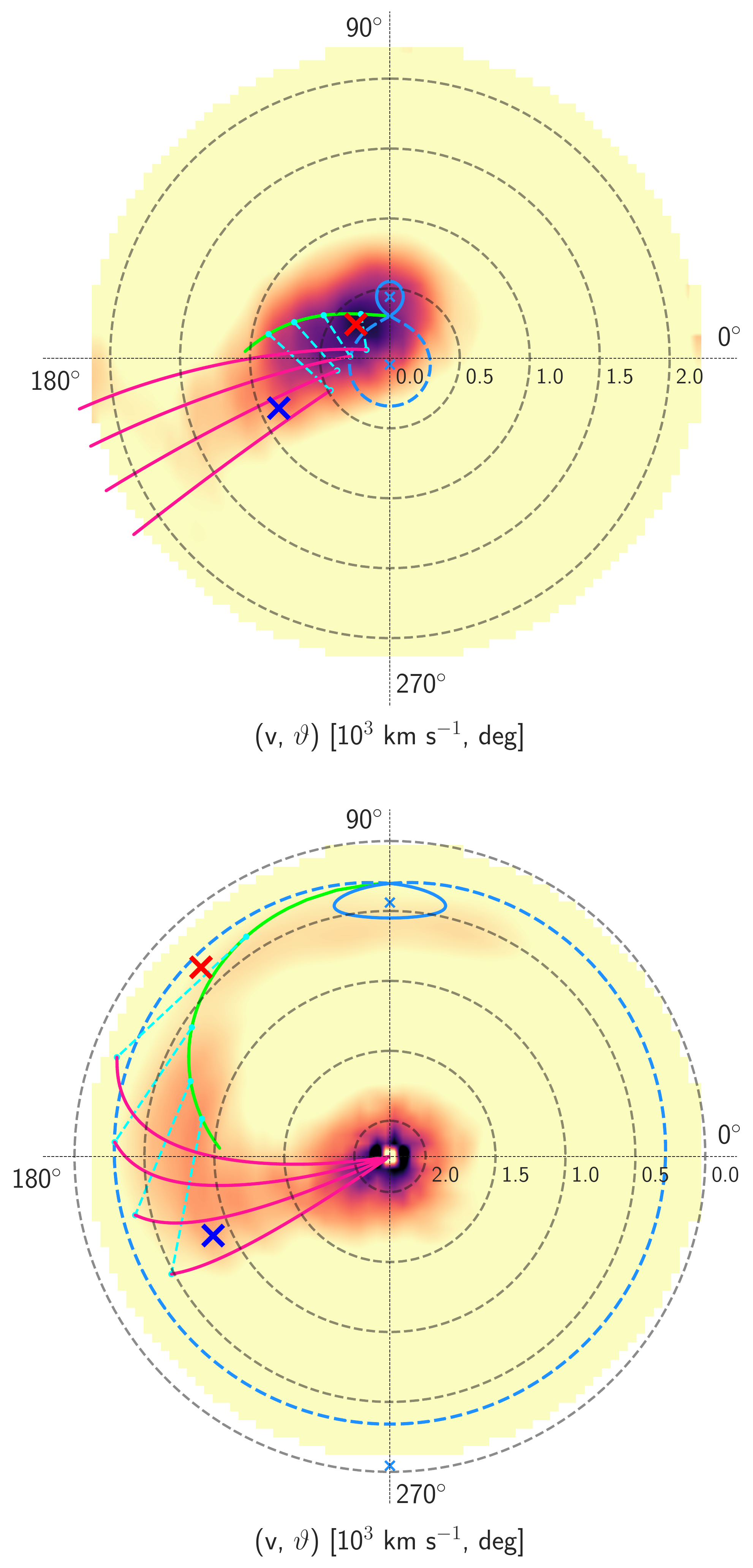}
   \end{minipage}%
   \caption{Doppler tomograms of the H$\alpha$ line are shown in the standard (top) and inside-out (bottom) projections, assuming that the narrow emission component originates on the donor's surface (left), and in the same configuration rotated by 45$^\circ$ (right). Overlaid on the maps is a binary model based on the derived system parameters, showing the Roche lobes of the white dwarf and the donor, as well as the ballistic and magnetic trajectories of the accretion flow in velocity space. The positions of the narrow (red cross) and broad (blue cross) components inferred from the modelling of the H$\alpha$ dynamical spectrum are also indicated.}
   \label{fig:doptomog_ha}
 \end{figure}

Doppler tomography provides additional information on the regions where emission lines are formed. This technique enables the reconstruction of the spatial distribution of emission-line sources in velocity space from phase-resolved spectroscopy. A detailed description of Doppler tomography can be found in \cite{Marsh2016, Kotze2015, Kotze2016}. To construct the Doppler tomograms, we used the code developed by \cite{Kotze2016}. The Doppler maps of the H$\alpha$ line obtained from the BTA phase-resolved spectra, shown in both the standard and inside-out projections, are presented in Fig.~\ref{fig:doptomog_ha}. The inside-out map differs from the standard one, as the velocity magnitude increases from the outer regions toward the center. This view is particularly useful for examining high-velocity areas of the accretion flow, which are often blurred or poorly resolved in standard maps (see \citealt{Kotze2015} for details). On the tomograms, we mark the locations of the bright narrow and faint broad emission components derived from the H$\alpha$ dynamic spectrum modelling. The figure shows two versions of the Doppler maps. The first was constructed using orbital phases calculated according to the ephemeris (\ref{eq:ephemeris}), assuming that the emission originates from the surface of the donor. The emission source appears elongated, suggesting that at least part of the emission arises from the accretion stream. In the second version, we applied a phase offset of $\Delta \varphi = 0.125$, resulting in a rotation of the tomograms by $\Delta \theta = 45^{\circ}$. This produces a Doppler map typical of polars, showing the gas flowing along the magnetically confined trajectory in the third quadrant or close to it. Accordingly, the narrow bright emission source lies close to the ballistic trajectory, possibly indicating that part of the emission originates within the ballistic section of the accretion flow rather than on the donor surface. In contrast, the broad faint source corresponds to the magnetic funnel region, characterized by high velocity dispersion. A similar accretion pattern was reported by \cite{Kolbin2023, Kochkina2023}, where the formation region of the narrow emission line component was interpreted as associated with the accretion stream. The dynamical spectra in Fig.~\ref{fig:dynspec_ha} further support an origin of the H$\alpha$ emission line in the accretion region. The flux shows a pronounced minimum around phase $\phi \approx 1.15$, likely caused by an eclipse of the emitting region. If the emission formed on the irradiated surface of the donor, the minimum would occur at phase $\phi = 1$.

\subsection{Magnetic field}
\label{sec:mag_field}

\begin{figure}[h!]
    \centering
    \includegraphics[width=0.7\linewidth]{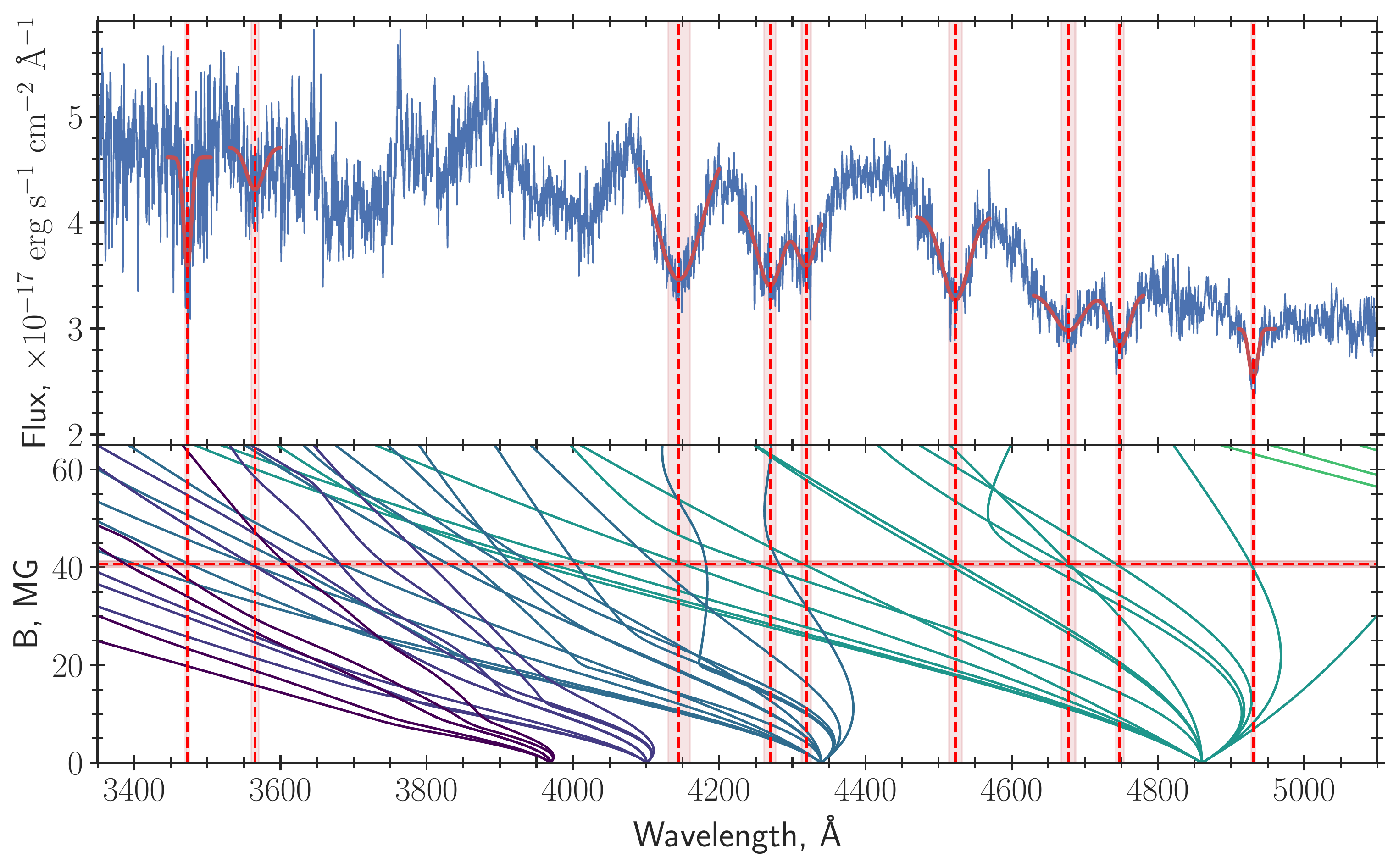}
    \caption{Top panel: Phase-averaged VLT spectrum showing Zeeman components. Best-fit Gaussian profiles and their centroids (red vertical lines) with widths of $\sim\sigma$ are overplotted. Bottom panel: Zeeman splitting diagram for the H$\beta$, H$\gamma$, H$\delta$, and H$\epsilon$ lines. The horizontal red line marks the derived mean magnetic field.}
    \label{fig:zeeman_vlt}
\end{figure}

An estimate of the white dwarf's magnetic field can be inferred from the Zeeman splitting of spectral lines. The VLT and BTA spectra show several Zeeman components of the Balmer lines in the blue spectral range of $3400 - 5000$~\AA. The individual spectra are too noisy to accurately determine the positions of the absorption features, so we used only phase-averaged spectra. The centroids of the components were determined by fitting their absorption profiles with Gaussian functions. The uncertainty in the central wavelength was adopted as the standard deviation $\sigma$ of the corresponding Gaussian fit. To compute the theoretical relation between the Zeeman component wavelengths and the magnetic field strength, we calculated the transition energies between hydrogen states in a strong magnetic field using the code of \cite{Schimeczek2014}. The mean magnetic field $B_\mathrm{wd}$ consistent with the observed splitting was then determined using a least-squares fit, and its uncertainty was evaluated via Monte Carlo simulations. The phase-averaged VLT UVB spectrum is shown in Fig.~\ref{fig:zeeman_vlt}, along with Gaussian fit profiles and their centroids representing the Zeeman-split features. A calculated diagram relating the Zeeman splitting to the magnetic field strength is also presented. In this way, we derived a mean magnetic field of $B = 40.7 \pm 0.5$~MG from the VLT data and $B = 40.3 \pm 1.2$~MG from the BTA data. As noted by \cite{Parsons2021} (who estimated 42~MG from cyclotron harmonics), determining the magnetic field via this method is affected by the broadening of Zeeman components caused by variations in the field strength across the white dwarf surface. We find that the magnetic field variations over the orbital period remain within the uncertainties.

\section{Cyclotron emission modelling}
\label{sec:cyclo_mod}

\subsection{Bombardment regime}
\label{sec:bomb_reg}

Based on two sets of phase-resolved spectroscopic data, we estimated the accretion parameters of {\obj} by modelling the cyclotron emission from the accretion spot. Since the system was in a low state during the observations, the theoretical spectra of the accretion spot were calculated using the so-called bombardment solution. In the presence of a strong magnetic field and a low accretion rate (for $\dot{m} (B/10^7~\text{G})^{-2.6} < 0.1~\text{g}~\text{s}^{-1}~\text{cm}^{-2}$; \citealt{Campbell2008}), the kinetic energy of the particles of the infalling ionized gas is released as cyclotron radiation over its stopping length in the atmosphere \citep{Kuijpers1982, Woelk1993}. In this case, no strong shock forms, and the temperature of the accretion spot is significantly lower than that expected for an adiabatic shock $T_\mathrm{sh} \sim 10-50$~keV. Following \cite{Woelk1993}, within this approximation the vertical profiles of the temperature $T$ and column density $x$ can be parameterized in terms of the local accretion rate $\dot{m}$ (in g~s$^{-1}$~cm$^{-2}$), magnetic field strength $B$, and white dwarf mass $M_\mathrm{wd}$. Under the high Faraday rotation prevailing in the accretion spot, the computation of the emergent cyclotron intensity reduces to solving the radiative transfer equations for the ordinary ($o$) and extraordinary ($e$) polarization modes:
\begin{equation}
    \begin{gathered}
        I_{o,e}(\omega, \theta, T) = \int^{x_s}_0 \frac{\alpha_{o,e}(\omega, \theta, T)}{2\rho \cos{\theta}} B(\omega, T) \times \exp\bigg(-\int^x_0 \frac{\alpha_{o,e}(\omega, \theta, T)}{\rho \cos{\theta}}d\tilde{x}\bigg)dx,
    \label{eq:transfer_solution}
    \end{gathered}
\end{equation}
where $T = T(x)$ and $\rho = \rho(x)$ denote the temperature and density profiles, $\theta$ is the angle between the magnetic field and the line of sight, $\alpha_{o,e}$ are the cyclotron absorption coefficients, and $B(\omega, T)$ is the Planck function. The integration is performed over the column mass $x$ along the depth of the accretion spot up to $x_s$. The absorption coefficients $\alpha_{o,e}$ were derived following the approach of \cite{Chanmugam1981}. The total cyclotron intensity is then expressed as $I = I_e + I_o$. A detailed description of the method for calculating both the spectrum and the vertical structure of the accretion spot within the bombardment solution is provided by \cite{Rousseau1996, Woelk1993}.

\subsection{Accretion spot model}
\label{sec:acc_spot}

To determine the accretion parameters of {\obj}, we employ a simplified model of a circular accretion spot that is homogeneous in the local accretion rate $\dot{m}$. The rotation axis of the white dwarf is inclined by an angle $i$ relative to the line of sight, while the magnetic pole is defined by a colatitude $\beta$ and an azimuth $\psi$, measured from the direction toward the donor. For convenience, we assume that the spot is located at the magnetic pole and that the magnetic field has a dipolar geometry. The spot size is specified by the angular radius $\alpha$, measured from the center of the white dwarf. The spot is divided into small segments of approximately equal area. To calculate the cyclotron emission intensity from an individual segment at a given rotational phase $\varphi$, it is necessary to determine the angle $\theta$ between the magnetic field direction and the line of sight. For this purpose, we use the equations from \cite{Cropper1989, Kolbin2020}.

\subsection{Accretion parameters}
\label{sec:acc_pars}

\begin{figure}[h!]
   \centering
   \includegraphics[width=\linewidth]{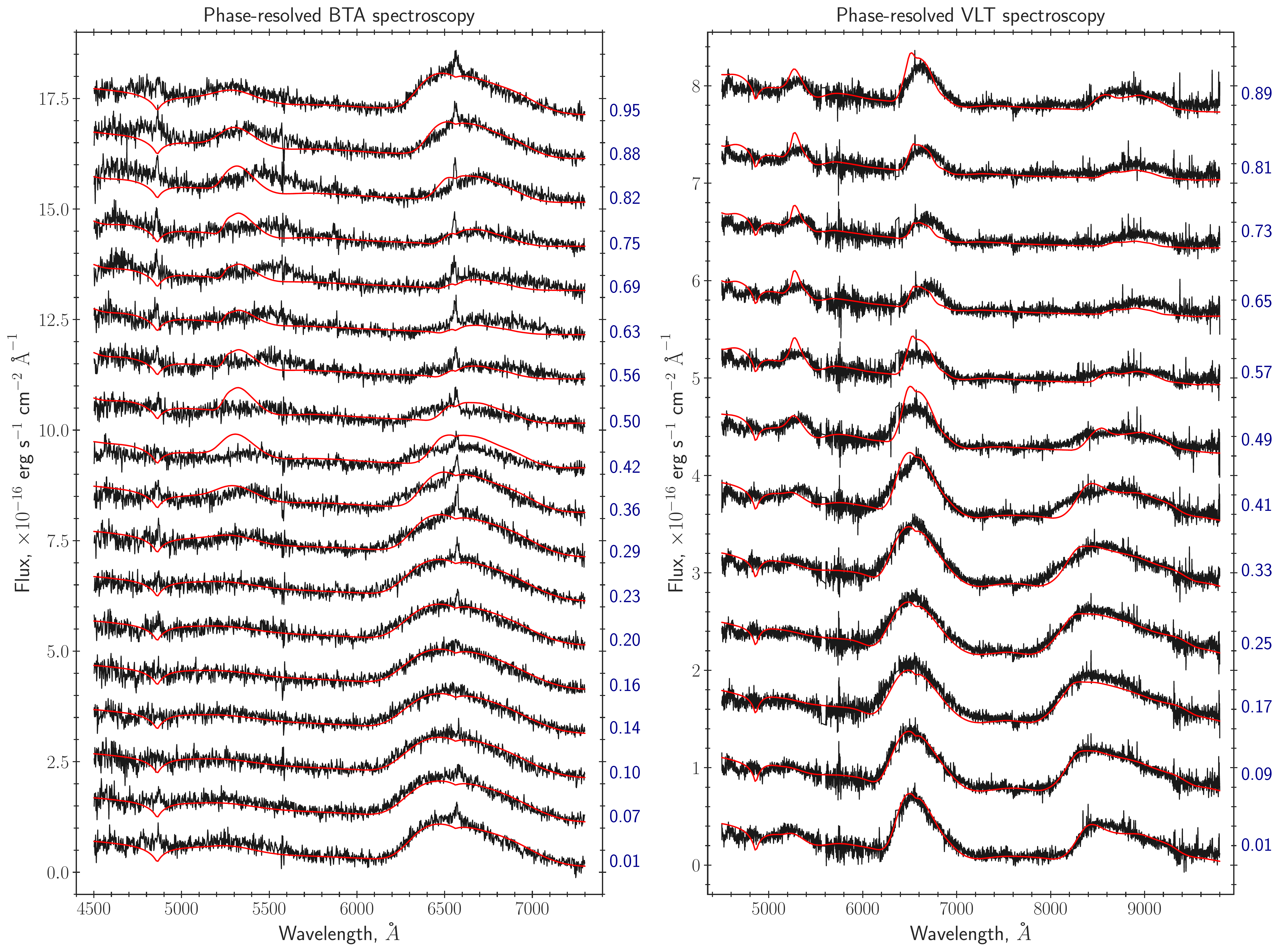}
   \caption{Modelling of the optical phase-resolved BTA (left) and VLT (right) spectroscopy using an orbitally modulated accretion spot. The synthetic spectra include a variable cyclotron emission and the white dwarf underlying flux. Orbital phases of the individual observations are indicated along the right-hand y-axis. For clarity, the spectra are vertically offset by a constant flux.}
   \label{fig:cycmod}
\end{figure}

The modelling of the phase-resolved spectroscopy was performed using a combination of variable emission from the accretion spot and a constant contribution from the white dwarf. The spot luminosity was calculated by integrating the emission of its individual surface elements, accounting for their visibility at each orbital phase. For the analysis of the VLT and BTA data, we adopted the white dwarf model obtained in Sect.~\ref{sec:sed}. The parameters of the accretion spot were determined through least-squares fitting, with the $\chi^2$ minimization performed using the Nelder--Mead algorithm. A comparison between the observed and synthetic spectra is presented in Fig.~\ref{fig:cycmod}. Overall, the model provides a good representation of both the spectral shape and the phase-dependent evolution of the cyclotron harmonics. The difficulty in reproducing the fifth cyclotron harmonic in the BTA data indicates that the single spot model, while adequate for the lower-order harmonic, cannot fully describe the phase-resolved spectra. In particular, the redshift of the fifth harmonic observed at phases $\phi = 0.4 - 0.8$ suggests the presence of an additional emitting region with a slightly lower magnetic field strength $\approx 39.5$~MG, possibly located near the opposite magnetic pole (the magnetic field strength cannot be reliably constrained due to the lack of clear signatures of other cyclotron harmonics from the second spot). A simple two-spot configuration is able to reproduce the position of the fifth harmonic for a subset of orbital phases without significantly affecting the fourth harmonic. The remaining discrepancy in the amplitude of the fifth harmonic in the BTA and VLT spectra is most likely related to the assumption of a homogeneous accretion spot. In the bombardment approximation, the relative intensities of cyclotron harmonics are highly sensitive to the local mass accretion rate, $\log \dot{m}$, with higher-order harmonics responding more strongly to its variations. Therefore, a more realistic description of the phase-dependent harmonic amplitudes would require a model with a locally inhomogeneous accretion rate, which is not considered in the present study.

Modelling of the phase-resolved spectral sets obtained with the BTA and VLT yielded the accretion spot parameters summarized in Table~\ref{tab:cyc_pars}. Parameter uncertainties were estimated using a Monte--Carlo method. We note that the parameters derived at the two different observing epochs are mutually consistent. In particular, closely matching values were determined for the inclination $i$, the magnetic colatitude $\beta$, and the magnetic field strength $B_\mathrm{m}$. The discrepancy in the azimuthal angle $\Delta \psi = 26 \pm 10^{\circ}$ may reflect the higher accretion rate during the BTA observations. The increased ram pressure of the accretion flow is expected to delay its magnetic coupling, resulting in a larger azimuthal longitude of the accretion spot. We also find that the magnetic field strength in the accretion region $B_\mathrm{m}$ slightly exceeds the value inferred from the Zeeman-splitting components. This is expected, as the cyclotron emitting region is located close to the magnetic pole, where the field strength is higher than the disk-averaged value over the visible surface of the white dwarf. Within the heated atmosphere approximation \citep{Woelk1993}, the maximum temperature reached in the vertically structured accretion spot is $T_\mathrm{max} = 1.67 \pm 0.17$~keV for the BTA data and $1.27 \pm 0.14$~keV for the VLT data. In both cases, the fractional area of the cyclotron source $f = S_\mathrm{spot} / S_\mathrm{wd}$ lies within the range $10^{-4} - 10^{-3}$, typical of polars \citep{Cropper1990}. In the low state, the accretion luminosity is assumed to be dominated by cyclotron radiation $L_\mathrm{acc} \approx L_\mathrm{cyc}$. Integrating the synthetic VLT spectrum, including all cyclotron harmonics, yields $L_\mathrm{cyc} = 2 \times 10^{30}$~erg~s$^{-1}$. The corresponding total accretion rate, derived from $L_\mathrm{acc} = G \dot{M} M_\mathrm{wd} / R_\mathrm{wd}$, where $G$ is the gravitational constant, is $\dot{M} = 2.4 \times 10^{-13}~M_{\odot}$~yr$^{-1}$, in agreement with the model results. The higher accretion rate during the BTA epoch relative to the VLT observations $\dot{M}_\mathrm{BTA} / \dot{M}_\mathrm{VLT} \sim 10^{-0.4\Delta m}$ is consistent with the increase in the optical brightness of the system $\Delta m \approx 0.5^\mathrm{m}$ observed in the long-term light curve. 

\begin{table}
   \begin{center}
   \caption[]{Parameters of the cyclotron emission region in {\obj}, derived from modelling the BTA and VLT phase-resolved spectroscopy within the bombardment approximation.}\label{tab:cyc_pars}
   \begin{tabular}{lccc}
   \hline\noalign{\smallskip}
   Parameters &  BTA spectra      & VLT spectra                    \\
   \hline\noalign{\smallskip}
   $i,~^\circ$                 & $69.6 \pm 2.8$                   & $68.5 \pm 3.1$                     \\
   $\beta,~^\circ$             & $16.4 \pm 7.4$                   & $16.4 \pm 6.6$                     \\
   $\psi,~^\circ$              & $50.6 \pm 9.2$                   & $76.4 \pm 7.5$                     \\
   $\log~\dot{m}$              & $-2.58 \pm 0.11$                 & $-2.86 \pm 0.11$                    \\ 
   $B_\mathrm{m},$~MG          & $40.91 \pm 0.16$                 & $41.14 \pm 0.13$                   \\
   $T_\mathrm{max},$~keV       & $1.67 \pm 0.17$                  & $1.27 \pm 0.14$                             \\
   $f,~\times 10^{-3}$         & $1.4 \pm 0.5$                    & $1.6 \pm 0.5$       \\
   $\dot{M},~\times 10^{-12}~M_{\odot}$~yr$^{-1}$     & $4.07 \pm 0.46$  & $2.50 \pm 0.31$  \\
   \noalign{\smallskip}\hline
   \end{tabular}
   \end{center}
\end{table}

\subsection{X-ray emission}
\label{sec:xray}

Between 2007 and 2018, {\obj} was observed with the Swift/XRT telescope while in a low state. The total exposure time was approximately $\sim 15000$~s; however, no more than 10 source counts were detected. Due to this limitation, it was difficult to model X-ray spectrum and to determine the X-ray flux with sufficient accuracy to assess its contribution to the accretion luminosity in the low state.

Using eROSITA observations in the soft X-ray band (0.2-2.3~keV), \cite{Giraldo2024} reported a flux of $F_\mathrm{x} = (6.71 \pm 0.55) \times 10^{-13}$~erg~s$^{-1}$~cm$^{-2}$ for {\obj}. The authors used the merged eRASS:3 catalog, which includes three surveys conducted between December 2019 and June 2021. According to the optical light curve shown in Fig.~\ref{fig:longterm_lc}, the source was in an intermediate state during this period. The accretion rate estimated by \cite{Giraldo2024} from the X-ray luminosity $L_\mathrm{x} = 4 \times 10^{31}$~erg~s$^{-1}$ was $\dot{M} = 2.3 \times 10^{-12}~M_\odot$~yr$^{-1}$. However, this value should be considered a lower limit, as the total accretion rate also includes cyclotron cooling in addition to X-ray bremsstrahlung. If the increase in optical brightness by $\sim 1.5 - 2^\mathrm{m}$ during the transition from the low to intermediate state is primarily due to enhanced cyclotron emission, the corresponding cyclotron luminosity $L_\mathrm{cyc}$ is expected to rise to $\sim (0.8-1.3) \times 10^{31}$~erg~s$^{-1}$. Accounting for both cooling mechanisms in the accretion spot $L_\mathrm{acc} = L_\mathrm{x} + L_\mathrm{cyc}$ the accretion rate in the intermediate state reaches $\dot{M} \sim 6 \times 10^{-12}~M_\odot$~yr$^{-1}$, roughly an order of magnitude higher than in the low state.

\section{Conslusion}
\label{sec:concl}

In this work, we performed a comprehensive analysis of the photometric and spectroscopic observations of the polar {\obj} in its low state. By combining data from multiple photometric surveys, we obtained a long-term light curve spanning approximately 20 years. During this period, the system transitioned into a high state at least three times, brightening by $\sim 2 - 3^{\rm m}$. Based on data from PTF, ZTF, and GALEX, we refined the orbital period and established a correlation between the optical and ultraviolet fluxes. This correlation indicates that the ultraviolet emission originates in the vicinity of the accretion spot. Through modelling of the spectral energy distribution in the optical and ultraviolet bands, we estimated the atmospheric parameters of the white dwarf to be $T_\mathrm{eff} = 12700 \pm 350$~K, $\log g = 8.2 \pm 0.4$, and a mass of $M_\mathrm{wd} = 0.74 \pm 0.05~M_\odot$. The phase-resolved spectroscopy of {\obj} obtained with the BTA and VLT telescopes corresponds to epochs when the system was in a low state. However, a detailed inspection of the BTA spectra revealed the presence of the Balmer emission lines H$\alpha$, H$\beta$, H$\gamma$, as well as HeI(5876), whereas all of them except H$\alpha$ are nearly absent in the VLT data. This suggests that the mass-transfer rate was lower at the time of the VLT observations. Analysis of the H$\alpha$ profile variability over the orbital period using BTA data shows that it can be described by a composite model consisting of narrow and broad components. Doppler tomography reveals the presence of an accretion structure. If most of the emission is assumed to originate from the irradiated surface of the donor star, the resulting accretion stream trajectory is poorly reproduced by the combination of its ballistic and magnetically confined parts. However, rotating the tomograms by $\approx 45^\circ$ aligns them with the expected positions of these structures. This implies that the H$\alpha$ emission is formed primarily within the accretion stream rather than on the donor surface. Analysis of the Zeeman splitting yields an average (disk-integrated) surface magnetic field strength of 40.5~MG for the white dwarf, with its variation over the spin period being less than 1~MG. Using the two sets of phase-resolved spectroscopy, we modelled the rotation-modulated cyclotron emission from the accretion spot. Both spectral series provided consistent estimates for the location of the emission region on the white dwarf surface. The total accretion rate during the BTA observations $\dot{M} = (4.07 \pm 0.46) \times 10^{-13}~M_\odot$~yr$^{-1}$ was, as expected, higher than that inferred from the VLT data $\dot{M} = (2.50 \pm 0.31) \times 10^{-13}~M_\odot$~yr$^{-1}$. The difference of $\sim 26^\circ$ in the azimuthal angle $\psi$ may reflect changes in the accretion rate affecting the location of the threading region and, consequently, the accretion spot coordinates. The modelled accretion rate is consistent with the value derived from the luminosity of the cyclotron harmonics, $\dot{M} = 2.4 \times 10^{-13}~M_\odot$~yr$^{-1}$, and is similar to the estimate reported in \cite{Schmidt2007}.

\normalem
\begin{acknowledgements}

We are grateful to Arkadiy Sarkisyan for performing spectral observations on the BTA telescope.
Observations with the SAO RAS telescopes are supported by the Ministry of Science and Higher Education of the Russian Federation. The modernization of the observatory's equipment is carried out under the national project "Science and universities".

\end{acknowledgements}
  
\bibliographystyle{raa}
\bibliography{bibtex}

@ARTICLE{Afanasiev2005,
       author = {{Afanasiev}, V.~L. and {Moiseev}, A.~V.},
        title = "{The SCORPIO Universal Focal Reducer of the 6-m Telescope}",
      journal = {Astronomy Letters},
         year = 2005,
        month = mar,
       volume = {31},
       number = {3},
        pages = {194-204},
          doi = {10.1134/1.1883351},
archivePrefix = {arXiv},
       eprint = {astro-ph/0502095},
 primaryClass = {astro-ph},
       adsurl = {https://ui.adsabs.harvard.edu/abs/2005AstL...31..194A}
}

@ARTICLE{Beuermann1987,
       author = {{Beuermann}, K. and {Stella}, L. and {Patterson}, J.},
        title = "{Einstein Observations of EF Eridani (2A 0311-227), The Textbook Example of AM Herculis--Type Systems}",
      journal = {\apj},
         year = 1987,
        month = may,
       volume = {316},
        pages = {360},
          doi = {10.1086/165206},
       adsurl = {https://ui.adsabs.harvard.edu/abs/1987ApJ...316..360B}
}

@INPROCEEDINGS{Webbink2005,
       author = {{Webbink}, R.~F. and {Wickramasinghe}, D.~T.},
        title = "{A Model for Low Accretion Rate Polars}",
    booktitle = {The Astrophysics of Cataclysmic Variables and Related Objects},
         year = 2005,
       editor = {{Hameury}, J. -M. and {Lasota}, J. -P.},
       series = {Astronomical Society of the Pacific Conference Series},
       volume = {330},
        month = aug,
        pages = {137},
       adsurl = {https://ui.adsabs.harvard.edu/abs/2005ASPC..330..137W}
}

@ARTICLE{Woelk1993,
       author = {{Woelk}, U. and {Beuermann}, K.},
        title = "{Temperature structure of a particle -- heated magnetic atmosphere}",
      journal = {\aap},
         year = 1993,
        month = dec,
       volume = {280},
       number = {1},
        pages = {169-173},
       adsurl = {https://ui.adsabs.harvard.edu/abs/1993A&A...280..169W}
}

@ARTICLE{Green2018,
       author = {{Green}, Gregory M. and {Schlafly}, Edward F. and {Finkbeiner}, Douglas and {Rix}, Hans-Walter and {Martin}, Nicolas and {Burgett}, William and {Draper}, Peter W. and {Flewelling}, Heather and {Hodapp}, Klaus and {Kaiser}, Nicholas and {Kudritzki}, Rolf-Peter and {Magnier}, Eugene A. and {Metcalfe}, Nigel and {Tonry}, John L. and {Wainscoat}, Richard and {Waters}, Christopher},
        title = "{Galactic reddening in 3D from stellar photometry - an improved map}",
      journal = {\mnras},
         year = 2018,
        month = jul,
       volume = {478},
       number = {1},
        pages = {651-666},
          doi = {10.1093/mnras/sty1008},
archivePrefix = {arXiv},
       eprint = {1801.03555},
 primaryClass = {astro-ph.GA},
       adsurl = {https://ui.adsabs.harvard.edu/abs/2018MNRAS.478..651G}
}

@ARTICLE{Duffy2022,
       author = {{Duffy}, C. and {Ramsay}, G. and {Wu}, Kinwah and {Mason}, Paul A. and {Hakala}, P. and {Steeghs}, D. and {Wood}, M.~A.},
        title = "{Short-duration accretion states of Polars as seen in TESS and ZTF data}",
      journal = {\mnras},
         year = 2022,
        month = nov,
       volume = {516},
       number = {3},
        pages = {3144-3158},
          doi = {10.1093/mnras/stac2482},
archivePrefix = {arXiv},
       eprint = {2208.14855},
 primaryClass = {astro-ph.SR},
       adsurl = {https://ui.adsabs.harvard.edu/abs/2022MNRAS.516.3144D}
}

@ARTICLE{Debes2006,
       author = {{Debes}, J.~H. and {L{\'o}pez-Morales}, M. and {Bonanos}, A.~Z. and {Weinberger}, A.~J.},
        title = "{Detection of a Variable Infrared Excess around SDSS J121209.31+013627.7}",
      journal = {\apjl},
         year = 2006,
        month = aug,
       volume = {647},
       number = {2},
        pages = {L147-L150},
          doi = {10.1086/507486},
archivePrefix = {arXiv},
       eprint = {astro-ph/0607151},
 primaryClass = {astro-ph},
       adsurl = {https://ui.adsabs.harvard.edu/abs/2006ApJ...647L.147D}
}

@ARTICLE{Drake2009,
       author = {{Drake}, A.~J. and {Djorgovski}, S.~G. and {Mahabal}, A. and {Beshore}, E. and {Larson}, S. and {Graham}, M.~J. and {Williams}, R. and {Christensen}, E. and {Catelan}, M. and {Boattini}, A. and {Gibbs}, A. and {Hill}, R. and {Kowalski}, R.},
        title = "{First Results from the Catalina Real-Time Transient Survey}",
      journal = {\apj},
         year = 2009,
        month = may,
       volume = {696},
       number = {1},
        pages = {870-884},
          doi = {10.1088/0004-637X/696/1/870},
archivePrefix = {arXiv},
       eprint = {0809.1394},
 primaryClass = {astro-ph},
       adsurl = {https://ui.adsabs.harvard.edu/abs/2009ApJ...696..870D}
}

@ARTICLE{Kausch2015,
       author = {{Kausch}, W. and {Noll}, S. and {Smette}, A. and {Kimeswenger}, S. and {Barden}, M. and {Szyszka}, C. and {Jones}, A.~M. and {Sana}, H. and {Horst}, H. and {Kerber}, F.},
        title = "{Molecfit: A general tool for telluric absorption correction. II. Quantitative evaluation on ESO-VLT/X-Shooterspectra}",
      journal = {\aap},
         year = 2015,
        month = apr,
       volume = {576},
          eid = {A78},
        pages = {A78},
          doi = {10.1051/0004-6361/201423909},
archivePrefix = {arXiv},
       eprint = {1501.07265},
 primaryClass = {astro-ph.IM},
       adsurl = {https://ui.adsabs.harvard.edu/abs/2015A&A...576A..78K}
}

@ARTICLE{Koester2010,
       author = {{Koester}, D.},
        title = "{White dwarf spectra and atmosphere models}",
      journal = {\memsai},
         year = 2010,
        month = jan,
       volume = {81},
        pages = {921-931},
       adsurl = {https://ui.adsabs.harvard.edu/abs/2010MmSAI..81..921K}
}

@ARTICLE{King1998,
       author = {{King}, Andrew R. and {Cannizzo}, John K.},
        title = "{Low States in Cataclysmic Variables}",
      journal = {\apj},
         year = 1998,
        month = may,
       volume = {499},
       number = {1},
        pages = {348-354},
          doi = {10.1086/305630},
       adsurl = {https://ui.adsabs.harvard.edu/abs/1998ApJ...499..348K}
}

@ARTICLE{Knigge2011,
       author = {{Knigge}, Christian and {Baraffe}, Isabelle and {Patterson}, Joseph},
        title = "{The Evolution of Cataclysmic Variables as Revealed by Their Donor Stars}",
      journal = {\apjs},
         year = 2011,
        month = jun,
       volume = {194},
       number = {2},
          eid = {28},
        pages = {28},
          doi = {10.1088/0067-0049/194/2/28},
archivePrefix = {arXiv},
       eprint = {1102.2440},
 primaryClass = {astro-ph.SR},
       adsurl = {https://ui.adsabs.harvard.edu/abs/2011ApJS..194...28K}
}

@ARTICLE{Kolbin2023,
       author = {{Kolbin}, A.~I. and {Borisov}, N.~V. and {Burenkov}, A.~N. and {Spiridonova}, O.~I. and {Bikmaev}, I.~F. and {Suslikov}, M.~V.},
        title = "{Optical Study of the Polar BM CrB in a Low State}",
      journal = {Astronomy Letters},
         year = 2023,
        month = mar,
       volume = {49},
       number = {3},
        pages = {129-140},
          doi = {10.1134/S1063773723030040},
archivePrefix = {arXiv},
       eprint = {2304.13655},
 primaryClass = {astro-ph.HE},
       adsurl = {https://ui.adsabs.harvard.edu/abs/2023AstL...49..129K}
}

@ARTICLE{Kochkina2023,
       author = {{Kochkina}, V. Yu. and {Kolbin}, A.~I. and {Borisov}, N.~V. and {Bikmaev}, I.~F.},
        title = "{Nature of the Eclipsing Polar 1RXS J184542.4+483134}",
      journal = {Astronomy Letters},
         year = 2023,
        month = dec,
       volume = {49},
       number = {11},
        pages = {706-721},
          doi = {10.1134/S1063773723110051},
       adsurl = {https://ui.adsabs.harvard.edu/abs/2023AstL...49..706K}
}

@ARTICLE{Vallenari2023,
       author = {{Gaia Collaboration} and {Vallenari}, A. and {Brown}, A.~G.~A. and {Prusti}, T. and {de Bruijne}, J.~H.~J. and {Arenou}, F. and {Babusiaux}, C. and {Biermann}, M. and {Creevey}, O.~L. and {Ducourant}, C. and {Evans}, D.~W. and {Eyer}, L. and {Guerra}, R. and {Hutton}, A. and {Jordi}, C. and {Klioner}, S.~A. and {Lammers}, U.~L. and {Lindegren}, L. and {Luri}, X. and {Mignard}, F. and {Panem}, C. and {Pourbaix}, D. and {Randich}, S. and {Sartoretti}, P. and {Soubiran}, C. and {Tanga}, P. and {Walton}, N.~A. and {Bailer-Jones}, C.~A.~L. and {Bastian}, U. and {Drimmel}, R. and {Jansen}, F. and {Katz}, D. and {Lattanzi}, M.~G. and {van Leeuwen}, F. and {Bakker}, J. and {Cacciari}, C. and {Casta{\~n}eda}, J. and {De Angeli}, F. and {Fabricius}, C. and {Fouesneau}, M. and {Fr{\'e}mat}, Y. and {Galluccio}, L. and {Guerrier}, A. and {Heiter}, U. and {Masana}, E. and {Messineo}, R. and {Mowlavi}, N. and {Nicolas}, C. and {Nienartowicz}, K. and {Pailler}, F. and {Panuzzo}, P. and {Riclet}, F. and {Roux}, W. and {Seabroke}, G.~M. and {Sordo}, R. and {Th{\'e}venin}, F. and {Gracia-Abril}, G. and {Portell}, J. and {Teyssier}, D. and {Altmann}, M. and {Andrae}, R. and {Audard}, M. and {Bellas-Velidis}, I. and {Benson}, K. and {Berthier}, J. and {Blomme}, R. and {Burgess}, P.~W. and {Busonero}, D. and {Busso}, G. and {C{\'a}novas}, H. and {Carry}, B. and {Cellino}, A. and {Cheek}, N. and {Clementini}, G. and {Damerdji}, Y. and {Davidson}, M. and {de Teodoro}, P. and {Nu{\~n}ez Campos}, M. and {Delchambre}, L. and {Dell'Oro}, A. and {Esquej}, P. and {Fern{\'a}ndez-Hern{\'a}ndez}, J. and {Fraile}, E. and {Garabato}, D. and {Garc{\'\i}a-Lario}, P. and {Gosset}, E. and {Haigron}, R. and {Halbwachs}, J.-L. and {Hambly}, N.~C. and {Harrison}, D.~L. and {Hern{\'a}ndez}, J. and {Hestroffer}, D. and {Hodgkin}, S.~T. and {Holl}, B. and {Jan{\ss}en}, K. and {Jevardat de Fombelle}, G. and {Jordan}, S. and {Krone-Martins}, A. and {Lanzafame}, A.~C. and {L{\"o}ffler}, W. and {Marchal}, O. and {Marrese}, P.~M. and {Moitinho}, A. and {Muinonen}, K. and {Osborne}, P. and {Pancino}, E. and {Pauwels}, T. and {Recio-Blanco}, A. and {Reyl{\'e}}, C. and {Riello}, M. and {Rimoldini}, L. and {Roegiers}, T. and {Rybizki}, J. and {Sarro}, L.~M. and {Siopis}, C. and {Smith}, M. and {Sozzetti}, A. and {Utrilla}, E. and {van Leeuwen}, M. and {Abbas}, U. and {{\'A}brah{\'a}m}, P. and {Abreu Aramburu}, A. and {Aerts}, C. and {Aguado}, J.~J. and {Ajaj}, M. and {Aldea-Montero}, F. and {Altavilla}, G. and {{\'A}lvarez}, M.~A. and {Alves}, J. and {Anders}, F. and {Anderson}, R.~I. and {Anglada Varela}, E. and {Antoja}, T. and {Baines}, D. and {Baker}, S.~G. and {Balaguer-N{\'u}{\~n}ez}, L. and {Balbinot}, E. and {Balog}, Z. and {Barache}, C. and {Barbato}, D. and {Barros}, M. and {Barstow}, M.~A. and {Bartolom{\'e}}, S. and {Bassilana}, J.-L. and {Bauchet}, N. and {Becciani}, U. and {Bellazzini}, M. and {Berihuete}, A. and {Bernet}, M. and {Bertone}, S. and {Bianchi}, L. and {Binnenfeld}, A. and {Blanco-Cuaresma}, S. and {Blazere}, A. and {Boch}, T. and {Bombrun}, A. and {Bossini}, D. and {Bouquillon}, S. and {Bragaglia}, A. and {Bramante}, L. and {Breedt}, E. and {Bressan}, A. and {Brouillet}, N. and {Brugaletta}, E. and {Bucciarelli}, B. and {Burlacu}, A. and {Butkevich}, A.~G. and {Buzzi}, R. and {Caffau}, E. and {Cancelliere}, R. and {Cantat-Gaudin}, T. and {Carballo}, R. and {Carlucci}, T. and {Carnerero}, M.~I. and {Carrasco}, J.~M. and {Casamiquela}, L. and {Castellani}, M. and {Castro-Ginard}, A. and {Chaoul}, L. and {Charlot}, P. and {Chemin}, L. and {Chiaramida}, V. and {Chiavassa}, A. and {Chornay}, N. and {Comoretto}, G. and {Contursi}, G. and {Cooper}, W.~J. and {Cornez}, T. and {Cowell}, S. and {Crifo}, F. and {Cropper}, M. and {Crosta}, M. and {Crowley}, C. and {Dafonte}, C. and {Dapergolas}, A. and {David}, M. and {David}, P. and {de Laverny}, P. and {De Luise}, F. and {De March}, R.},
        title = "{Gaia Data Release 3. Summary of the content and survey properties}",
      journal = {\aap},
         year = 2023,
        month = jun,
       volume = {674},
          eid = {A1},
        pages = {A1},
          doi = {10.1051/0004-6361/202243940},
archivePrefix = {arXiv},
       eprint = {2208.00211},
 primaryClass = {astro-ph.GA},
       adsurl = {https://ui.adsabs.harvard.edu/abs/2023A&A...674A...1G}
}

@ARTICLE{Kotze2015,
       author = {{Kotze}, E.~J. and {Potter}, S.~B. and {McBride}, V.~A.},
        title = "{Exploring inside-out Doppler tomography: non-magnetic cataclysmic variables}",
      journal = {\aap},
         year = 2015,
        month = jul,
       volume = {579},
          eid = {A77},
        pages = {A77},
          doi = {10.1051/0004-6361/201526381},
archivePrefix = {arXiv},
       eprint = {1507.05213},
 primaryClass = {astro-ph.SR},
       adsurl = {https://ui.adsabs.harvard.edu/abs/2015A&A...579A..77K}
}

@ARTICLE{Kotze2016,
       author = {{Kotze}, E.~J. and {Potter}, S.~B. and {McBride}, V.~A.},
        title = "{Exploring inside-out Doppler tomography: magnetic cataclysmic variables}",
      journal = {\aap},
         year = 2016,
        month = oct,
       volume = {595},
          eid = {A47},
        pages = {A47},
          doi = {10.1051/0004-6361/201629120},
archivePrefix = {arXiv},
       eprint = {1610.09841},
 primaryClass = {astro-ph.SR},
       adsurl = {https://ui.adsabs.harvard.edu/abs/2016A&A...595A..47K}
}

@ARTICLE{Cropper1989,
       author = {{Cropper}, Mark},
        title = "{The accretion region in AM HER systems : insights from polarimetry ofV834 Cen.}",
      journal = {\mnras},
         year = 1989,
        month = feb,
       volume = {236},
        pages = {935-957},
          doi = {10.1093/mnras/236.4.935},
       adsurl = {https://ui.adsabs.harvard.edu/abs/1989MNRAS.236..935C}
}

@ARTICLE{Cropper1990,
       author = {{Cropper}, Mark},
        title = "{The Polars}",
      journal = {\ssr},
         year = 1990,
        month = dec,
       volume = {54},
       number = {3-4},
        pages = {195-295},
          doi = {10.1007/BF00177799},
       adsurl = {https://ui.adsabs.harvard.edu/abs/1990SSRv...54..195C}
}

@PHDTHESIS{Campbell2008,
       author = {{Campbell}, Ryan},
        title = "{Phase-resolved cyclotron spectroscopy of polars}",
       school = {New Mexico State University},
         year = 2008,
        month = jan,
       adsurl = {https://ui.adsabs.harvard.edu/abs/2008PhDT........27C}
}

@ARTICLE{Campbell2008_3,
       author = {{Campbell}, Ryan K. and {Harrison}, Thomas E. and {Kafka}, Stella},
        title = "{Cyclotron Modeling Phase-Resolved Infrared Spectroscopy of Polars. III. AM Herculis and ST Leo Minoris}",
      journal = {\apj},
         year = 2008,
        month = aug,
       volume = {683},
       number = {1},
        pages = {409-423},
          doi = {10.1086/589179},
       adsurl = {https://ui.adsabs.harvard.edu/abs/2008ApJ...683..409C}
}

@ARTICLE{Livio94,
       author = {{Livio}, M. and {Pringle}, J.~E.},
        title = "{Star Spots and the Period Gap in Cataclysmic Variables}",
      journal = {\apj},
         year = 1994,
        month = jun,
       volume = {427},
        pages = {956},
          doi = {10.1086/174202},
       adsurl = {https://ui.adsabs.harvard.edu/abs/1994ApJ...427..956L}
}

@ARTICLE{Linnell2010,
       author = {{Linnell}, Albert P. and {Szkody}, Paula and {Plotkin}, Richard M. and {Holtzman}, Jon and {Seibert}, Mark and {Harrison}, Thomas E. and {Howell}, Steve B.},
        title = "{GALEX and Optical Light Curves of WX LMi, SDSSJ103100.5+202832.2, and SDSSJ121209.31+013627.7}",
      journal = {\apj},
         year = 2010,
        month = apr,
       volume = {713},
       number = {2},
        pages = {1183-1191},
          doi = {10.1088/0004-637X/713/2/1183},
archivePrefix = {arXiv},
       eprint = {1003.2564},
 primaryClass = {astro-ph.SR},
       adsurl = {https://ui.adsabs.harvard.edu/abs/2010ApJ...713.1183L}
}

@INPROCEEDINGS{Marsh2016,
       author = {{Marsh}, Thomas R. and {Schwope}, Axel D.},
        title = "{Doppler Tomography}",
    booktitle = {Astronomy at High Angular Resolution},
         year = 2016,
       editor = {{Boffin}, Henri M.~J. and {Hussain}, Gaitee and {Berger}, Jean-Philippe and {Schmidtobreick}, Linda},
       series = {Astrophysics and Space Science Library},
       volume = {439},
        month = jan,
        pages = {195},
          doi = {10.1007/978-3-319-39739-9_11},
       adsurl = {https://ui.adsabs.harvard.edu/abs/2016ASSL..439..195M}
}

@ARTICLE{Masci2019,
       author = {{Masci}, Frank J. and {Laher}, Russ R. and {Rusholme}, Ben and {Shupe}, David L. and {Groom}, Steven and {Surace}, Jason and {Jackson}, Edward and {Monkewitz}, Serge and {Beck}, Ron and {Flynn}, David and {Terek}, Scott and {Landry}, Walter and {Hacopians}, Eugean and {Desai}, Vandana and {Howell}, Justin and {Brooke}, Tim and {Imel}, David and {Wachter}, Stefanie and {Ye}, Quan-Zhi and {Lin}, Hsing-Wen and {Cenko}, S. Bradley and {Cunningham}, Virginia and {Rebbapragada}, Umaa and {Bue}, Brian and {Miller}, Adam A. and {Mahabal}, Ashish and {Bellm}, Eric C. and {Patterson}, Maria T. and {Juri{\'c}}, Mario and {Golkhou}, V. Zach and {Ofek}, Eran O. and {Walters}, Richard and {Graham}, Matthew and {Kasliwal}, Mansi M. and {Dekany}, Richard G. and {Kupfer}, Thomas and {Burdge}, Kevin and {Cannella}, Christopher B. and {Barlow}, Tom and {Van Sistine}, Angela and {Giomi}, Matteo and {Fremling}, Christoffer and {Blagorodnova}, Nadejda and {Levitan}, David and {Riddle}, Reed and {Smith}, Roger M. and {Helou}, George and {Prince}, Thomas A. and {Kulkarni}, Shrinivas R.},
        title = "{The Zwicky Transient Facility: Data Processing, Products, and Archive}",
      journal = {\pasp},
         year = 2019,
        month = jan,
       volume = {131},
       number = {995},
        pages = {018003},
          doi = {10.1088/1538-3873/aae8ac},
archivePrefix = {arXiv},
       eprint = {1902.01872},
 primaryClass = {astro-ph.IM},
       adsurl = {https://ui.adsabs.harvard.edu/abs/2019PASP..131a8003M}
}

@ARTICLE{Million2016,
       author = {{Million}, Chase and {Fleming}, Scott W. and {Shiao}, Bernie and {Seibert}, Mark and {Loyd}, Parke and {Tucker}, Michael and {Smith}, Myron and {Thompson}, Randy and {White}, Richard L.},
        title = "{gPhoton: The GALEX Photon Data Archive}",
      journal = {\apj},
         year = 2016,
        month = dec,
       volume = {833},
       number = {2},
          eid = {292},
        pages = {292},
          doi = {10.3847/1538-4357/833/2/292},
archivePrefix = {arXiv},
       eprint = {1609.09492},
 primaryClass = {astro-ph.IM},
       adsurl = {https://ui.adsabs.harvard.edu/abs/2016ApJ...833..292M}
}

@ARTICLE{Giraldo2024,
       author = {{Mu{\~n}oz-Giraldo}, Daniela and {Stelzer}, Beate and {Schwope}, Axel},
        title = "{Cataclysmic variables around the period-bounce: An eROSITA-enhanced multiwavelength catalog}",
      journal = {\aap},
         year = 2024,
        month = jul,
       volume = {687},
          eid = {A305},
        pages = {A305},
          doi = {10.1051/0004-6361/202449358},
archivePrefix = {arXiv},
       eprint = {2401.17298},
 primaryClass = {astro-ph.SR},
       adsurl = {https://ui.adsabs.harvard.edu/abs/2024A&A...687A.305M}
}

@ARTICLE{Nauenberg72,
       author = {{Nauenberg}, Michael},
        title = "{Analytic Approximations to the Mass-Radius Relation and Energy of Zero-Temperature Stars}",
      journal = {\apj},
         year = 1972,
        month = jul,
       volume = {175},
        pages = {417},
          doi = {10.1086/151568},
       adsurl = {https://ui.adsabs.harvard.edu/abs/1972ApJ...175..417N}
}

@ARTICLE{Parsons2021,
       author = {{Parsons}, S.~G. and {G{\"a}nsicke}, B.~T. and {Schreiber}, M.~R. and {Marsh}, T.~R. and {Ashley}, R.~P. and {Breedt}, E. and {Littlefair}, S.~P. and {Meusinger}, H.},
        title = "{Magnetic white dwarfs in post-common-envelope binaries}",
      journal = {\mnras},
         year = 2021,
        month = apr,
       volume = {502},
       number = {3},
        pages = {4305-4327},
          doi = {10.1093/mnras/stab284},
archivePrefix = {arXiv},
       eprint = {2101.08792},
 primaryClass = {astro-ph.SR},
       adsurl = {https://ui.adsabs.harvard.edu/abs/2021MNRAS.502.4305P}
}

@ARTICLE{Rousseau1996,
       author = {{Rousseau}, T. and {Fischer}, A. and {Beuermann}, K. and {Woelk}, U.},
        title = "{Determination of mass flow rates in AM Herculis binaries. I. General method and application to UZ Fornacis.}",
      journal = {\aap},
         year = 1996,
        month = jun,
       volume = {310},
        pages = {526-532},
       adsurl = {https://ui.adsabs.harvard.edu/abs/1996A&A...310..526R}
}

@ARTICLE{vanRoestel2025,
       author = {{van Roestel}, J. and {Rodriguez}, A.~C. and {Szkody}, P. and {Brown}, A.~J. and {Caiazzo}, I. and {Drake}, A. and {El-Badry}, K. and {Prince}, T. and {Rich}, R.~M.~R. and {Neill}, J.~D. and {Vanderbosch}, Z. and {Bellm}, E.~C. and {Dekany}, R. and {Feinstein}, F. and {Graham}, M. and {Groom}, S.~L. and {Helou}, G. and {Kulkarni}, S.~R. and {du Laz}, T. and {Mahabal}, A. and {Sharma}, Y. and {Sollerman}, J. and {Wold}, A.},
        title = "{Cyclotron emitting magnetic white dwarfs in post common-envelope binaries discovered with the Zwicky Transient Facility}",
      journal = {\aap},
         year = 2025,
        month = apr,
       volume = {696},
          eid = {A242},
        pages = {A242},
          doi = {10.1051/0004-6361/202451945},
archivePrefix = {arXiv},
       eprint = {2412.15153},
 primaryClass = {astro-ph.SR},
       adsurl = {https://ui.adsabs.harvard.edu/abs/2025A&A...696A.242V}
}

@ARTICLE{Smette2015,
       author = {{Smette}, A. and {Sana}, H. and {Noll}, S. and {Horst}, H. and {Kausch}, W. and {Kimeswenger}, S. and {Barden}, M. and {Szyszka}, C. and {Jones}, A.~M. and {Gallenne}, A. and {Vinther}, J. and {Ballester}, P. and {Taylor}, J.},
        title = "{Molecfit: A general tool for telluric absorption correction. I. Method and application to ESO instruments}",
      journal = {\aap},
         year = 2015,
        month = apr,
       volume = {576},
          eid = {A77},
        pages = {A77},
          doi = {10.1051/0004-6361/201423932},
archivePrefix = {arXiv},
       eprint = {1501.07239},
 primaryClass = {astro-ph.IM},
       adsurl = {https://ui.adsabs.harvard.edu/abs/2015A&A...576A..77S}
}

@ARTICLE{Stelzer2017,
       author = {{Stelzer}, B. and {de Martino}, D. and {Casewell}, S.~L. and {Wynn}, G.~A. and {Roy}, M.},
        title = "{X-ray orbital modulation of a white dwarf accreting from an L dwarf. The system SDSS J121209.31+013627.7}",
      journal = {\aap},
         year = 2017,
        month = feb,
       volume = {598},
          eid = {L6},
        pages = {L6},
          doi = {10.1051/0004-6361/201630038},
archivePrefix = {arXiv},
       eprint = {1701.02894},
 primaryClass = {astro-ph.SR},
       adsurl = {https://ui.adsabs.harvard.edu/abs/2017A&A...598L...6S}
}

@ARTICLE{Suslikov2025a,
       author = {{Suslikov}, M.~V. and {Kolbin}, A.~I. and {Borisov}, N.~V.},
        title = "{On Accretion in the Polar V379 Vir}",
      journal = {Astronomy Letters},
         year = 2025,
        month = feb,
       volume = {51},
       number = {2},
        pages = {79-88},
          doi = {10.1134/S1063773725700239},
archivePrefix = {arXiv},
       eprint = {2506.17674},
 primaryClass = {astro-ph.SR},
       adsurl = {https://ui.adsabs.harvard.edu/abs/2025AstL...51...79S}
}

@ARTICLE{Suslikov2025b,
       author = {{Suslikov}, M.~V. and {Kolbin}, A.~I. and {Borisov}, N.~V.},
        title = "{Phase-Resolved Spectroscopy of the Polar V379 Vir with a Brown Dwarf}",
      journal = {arXiv e-prints},
         year = 2025,
        month = nov,
          eid = {arXiv:2511.14338},
        pages = {arXiv:2511.14338},
          doi = {10.48550/arXiv.2511.14338},
archivePrefix = {arXiv},
       eprint = {2511.14338},
 primaryClass = {astro-ph.HE},
       adsurl = {https://ui.adsabs.harvard.edu/abs/2025arXiv251114338S}
}

@BOOK{Warner1995,
       author = {{Warner}, Brian},
        title = "{Cataclysmic variable stars}",
         year = 1995,
       volume = {28},
       adsurl = {https://ui.adsabs.harvard.edu/abs/1995cvs..book.....W}
}

@ARTICLE{Farihi2008,
       author = {{Farihi}, J. and {Burleigh}, M.~R. and {Hoard}, D.~W.},
        title = "{A Near-Infrared Spectroscopic Study of the Accreting Magnetic White Dwarf SDSS J121209.31+013627.7 and Its Substellar Companion}",
      journal = {\apj},
         year = 2008,
        month = feb,
       volume = {674},
       number = {1},
        pages = {421-430},
          doi = {10.1086/524933},
archivePrefix = {arXiv},
       eprint = {0710.4969},
 primaryClass = {astro-ph},
       adsurl = {https://ui.adsabs.harvard.edu/abs/2008ApJ...674..421F}
}

@ARTICLE{Flewelling2020,
       author = {{Flewelling}, H.~A. and {Magnier}, E.~A. and {Chambers}, K.~C. and {Heasley}, J.~N. and {Holmberg}, C. and {Huber}, M.~E. and {Sweeney}, W. and {Waters}, C.~Z. and {Calamida}, A. and {Casertano}, S. and {Chen}, X. and {Farrow}, D. and {Hasinger}, G. and {Henderson}, R. and {Long}, K.~S. and {Metcalfe}, N. and {Narayan}, G. and {Nieto-Santisteban}, M.~A. and {Norberg}, P. and {Rest}, A. and {Saglia}, R.~P. and {Szalay}, A. and {Thakar}, A.~R. and {Tonry}, J.~L. and {Valenti}, J. and {Werner}, S. and {White}, R. and {Denneau}, L. and {Draper}, P.~W. and {Hodapp}, K.~W. and {Jedicke}, R. and {Kaiser}, N. and {Kudritzki}, R.~P. and {Price}, P.~A. and {Wainscoat}, R.~J. and {Chastel}, S. and {McLean}, B. and {Postman}, M. and {Shiao}, B.},
        title = "{The Pan-STARRS1 Database and Data Products}",
      journal = {\apjs},
         year = 2020,
        month = nov,
       volume = {251},
       number = {1},
          eid = {7},
        pages = {7},
          doi = {10.3847/1538-4365/abb82d},
archivePrefix = {arXiv},
       eprint = {1612.05243},
 primaryClass = {astro-ph.IM},
       adsurl = {https://ui.adsabs.harvard.edu/abs/2020ApJS..251....7F}
}

@ARTICLE{Fitzpatrick1999,
       author = {{Fitzpatrick}, Edward L.},
        title = "{Correcting for the Effects of Interstellar Extinction}",
      journal = {\pasp},
         year = 1999,
        month = jan,
       volume = {111},
       number = {755},
        pages = {63-75},
          doi = {10.1086/316293},
archivePrefix = {arXiv},
       eprint = {astro-ph/9809387},
 primaryClass = {astro-ph},
       adsurl = {https://ui.adsabs.harvard.edu/abs/1999PASP..111...63F}
}

@BOOK{Hellier2001,
       author = {{Hellier}, Coel},
        title = "{Cataclysmic Variable Stars}",
         year = 2001,
       adsurl = {https://ui.adsabs.harvard.edu/abs/2001cvs..book.....H}
}

@ARTICLE{Hessman2000,
       author = {{Hessman}, F.~V. and {G{\"a}nsicke}, B.~T. and {Mattei}, J.~A.},
        title = "{The history and source of mass-transfer variations in AM Herculis}",
      journal = {\aap},
         year = 2000,
        month = sep,
       volume = {361},
        pages = {952-958},
       adsurl = {https://ui.adsabs.harvard.edu/abs/2000A&A...361..952H}
}

@ARTICLE{Horne1986,
       author = {{Horne}, K.},
        title = "{An optimal extraction algorithm for CCD spectroscopy.}",
      journal = {\pasp},
         year = 1986,
        month = jun,
       volume = {98},
        pages = {609-617},
          doi = {10.1086/131801},
       adsurl = {https://ui.adsabs.harvard.edu/abs/1986PASP...98..609H}
}

@ARTICLE{Chanmugam1981,
       author = {{Chanmugam}, G. and {Dulk}, G.~A.},
        title = "{Polarized radiation from hot plasmas and applications to AM HER binaries.}",
      journal = {\apj},
         year = 1981,
        month = mar,
       volume = {244},
        pages = {569-578},
          doi = {10.1086/158736},
       adsurl = {https://ui.adsabs.harvard.edu/abs/1981ApJ...244..569C}
}

@INPROCEEDINGS{Schwope2002,
       author = {{Schwope}, A.~D. and {Brunner}, H. and {Hambaryan}, V. and {Schwarz}, R.},
        title = "{LARPs -- Low-accretion rate polars}",
    booktitle = {The Physics of Cataclysmic Variables and Related Objects},
         year = 2002,
       editor = {{G{\"a}nsicke}, B.~T. and {Beuermann}, K. and {Reinsch}, K.},
       series = {Astronomical Society of the Pacific Conference Series},
       volume = {261},
        month = jan,
        pages = {102},
       adsurl = {https://ui.adsabs.harvard.edu/abs/2002ASPC..261..102S}
}

@ARTICLE{Schwope2007,
       author = {{Schwope}, A.~D. and {Staude}, A. and {Koester}, D. and {Vogel}, J.},
        title = "{XMM-Newton observations of EF Eridani: the textbook example of low-accretion rate polars}",
      journal = {\aap},
         year = 2007,
        month = jul,
       volume = {469},
       number = {3},
        pages = {1027-1031},
          doi = {10.1051/0004-6361:20066928},
archivePrefix = {arXiv},
       eprint = {astro-ph/0703561},
 primaryClass = {astro-ph},
       adsurl = {https://ui.adsabs.harvard.edu/abs/2007A&A...469.1027S}
}

@ARTICLE{Schwope2009,
       author = {{Schwope}, A.~D. and {Nebot Gomez-Moran}, A. and {Schreiber}, M.~R. and {G{\"a}nsicke}, B.~T.},
        title = "{Post common envelope binaries from the SDSS. VI. SDSS J120615.73+510047.0: a new low accretion rate magnetic binary}",
      journal = {\aap},
         year = 2009,
        month = jun,
       volume = {500},
       number = {2},
        pages = {867-872},
          doi = {10.1051/0004-6361/200911699},
archivePrefix = {arXiv},
       eprint = {0903.5417},
 primaryClass = {astro-ph.SR},
       adsurl = {https://ui.adsabs.harvard.edu/abs/2009A&A...500..867S}
}

@ARTICLE{Schimeczek2014,
       author = {{Schimeczek}, C. and {Wunner}, G.},
        title = "{Atomic Data for the Spectral Analysis of Magnetic DA White Dwarfs in the SDSS}",
      journal = {\apjs},
         year = 2014,
        month = jun,
       volume = {212},
       number = {2},
          eid = {26},
        pages = {26},
          doi = {10.1088/0067-0049/212/2/26},
       adsurl = {https://ui.adsabs.harvard.edu/abs/2014ApJS..212...26S}
}

@ARTICLE{Schmidt2005,
       author = {{Schmidt}, Gary D. and {Szkody}, Paula and {Silvestri}, Nicole M. and {Cushing}, Michael C. and {Liebert}, James and {Smith}, Paul S.},
        title = "{Discovery of a Magnetic White Dwarf/Probable Brown Dwarf Short-Period Binary}",
      journal = {\apjl},
         year = 2005,
        month = sep,
       volume = {630},
       number = {2},
        pages = {L173-L176},
          doi = {10.1086/491702},
archivePrefix = {arXiv},
       eprint = {astro-ph/0508043},
 primaryClass = {astro-ph},
       adsurl = {https://ui.adsabs.harvard.edu/abs/2005ApJ...630L.173S}
}

@ARTICLE{Schmidt2007,
       author = {{Schmidt}, Gary D. and {Szkody}, Paula and {Henden}, Arne and {Anderson}, Scott F. and {Lamb}, Don Q. and {Margon}, Bruce and {Schneider}, Donald P.},
        title = "{Two Additions to the New Class of Low Accretion Rate Magnetic Binaries}",
      journal = {\apj},
         year = 2007,
        month = jan,
       volume = {654},
       number = {1},
        pages = {521-526},
          doi = {10.1086/509613},
archivePrefix = {arXiv},
       eprint = {astro-ph/0610818},
 primaryClass = {astro-ph},
       adsurl = {https://ui.adsabs.harvard.edu/abs/2007ApJ...654..521S}
}

@ARTICLE{Schreiber2023,
       author = {{Schreiber}, Matthias R. and {Belloni}, Diogo and {van Roestel}, Jan},
        title = "{Period bouncers as detached magnetic cataclysmic variables}",
      journal = {\aap},
         year = 2023,
        month = nov,
       volume = {679},
          eid = {L8},
        pages = {L8},
          doi = {10.1051/0004-6361/202347766},
archivePrefix = {arXiv},
       eprint = {2310.17276},
 primaryClass = {astro-ph.SR},
       adsurl = {https://ui.adsabs.harvard.edu/abs/2023A&A...679L...8S}
}

@ARTICLE{Low2009,
       author = {{Law}, Nicholas M. and {Kulkarni}, Shrinivas R. and {Dekany}, Richard G. and {Ofek}, Eran O. and {Quimby}, Robert M. and {Nugent}, Peter E. and {Surace}, Jason and {Grillmair}, Carl C. and {Bloom}, Joshua S. and {Kasliwal}, Mansi M. and {Bildsten}, Lars and {Brown}, Tim and {Cenko}, S. Bradley and {Ciardi}, David and {Croner}, Ernest and {Djorgovski}, S. George and {van Eyken}, Julian and {Filippenko}, Alexei V. and {Fox}, Derek B. and {Gal-Yam}, Avishay and {Hale}, David and {Hamam}, Nouhad and {Helou}, George and {Henning}, John and {Howell}, D. Andrew and {Jacobsen}, Janet and {Laher}, Russ and {Mattingly}, Sean and {McKenna}, Dan and {Pickles}, Andrew and {Poznanski}, Dovi and {Rahmer}, Gustavo and {Rau}, Arne and {Rosing}, Wayne and {Shara}, Michael and {Smith}, Roger and {Starr}, Dan and {Sullivan}, Mark and {Velur}, Viswa and {Walters}, Richard and {Zolkower}, Jeff},
        title = "{The Palomar Transient Factory: System Overview, Performance, and First Results}",
      journal = {\pasp},
         year = 2009,
        month = dec,
       volume = {121},
       number = {886},
        pages = {1395},
          doi = {10.1086/648598},
archivePrefix = {arXiv},
       eprint = {0906.5350},
 primaryClass = {astro-ph.IM},
       adsurl = {https://ui.adsabs.harvard.edu/abs/2009PASP..121.1395L}
}

@article{Gao2012,
    author = "Gao, Fuchang and Han, Lixing",
    title = "{Implementing the Nelder-Mead simplex algorithm with~adaptive parameters}",
    doi = "10.1007/s10589-010-9329-3",
    journal = "Comput. Optim. Appl.",
    volume = "51",
    number = "1",
    pages = "259--277",
    year = "2012"
}

@ARTICLE{Lin2025,
       author = {{Lin}, Jiamao and {Ren}, Liangliang and {Li}, Chengyuan and {Nancy}, Elias-Rosa and {Cang}, Tianqi and {Ge}, Hongwei and {Thomas Tam}, Pak-Hin and {Huang}, Wenjun and {Li}, Yilong and {Wang}, Xiaofeng and {Huang}, Yang and {Ma}, Bo},
        title = "{Discovery and characterization of ZTF J0112+5827: An 80.9-minute polar with strong cyclotron features}",
      journal = {\aap},
         year = 2025,
        month = feb,
       volume = {694},
          eid = {A112},
        pages = {A112},
          doi = {10.1051/0004-6361/202452177},
archivePrefix = {arXiv},
       eprint = {2502.16059},
 primaryClass = {astro-ph.SR},
       adsurl = {https://ui.adsabs.harvard.edu/abs/2025A&A...694A.112L}
}

@ARTICLE{Liu2023,
       author = {{Liu}, Yiqi and {Hwang}, Hsiang-Chih and {Zakamska}, Nadia L. and {Thorstensen}, John R.},
        title = "{CSS1603+19: a low-mass polar near the cataclysmic variable period minimum}",
      journal = {\mnras},
         year = 2023,
        month = jun,
       volume = {522},
       number = {2},
        pages = {2719-2731},
          doi = {10.1093/mnras/stad1156},
archivePrefix = {arXiv},
       eprint = {2211.14945},
 primaryClass = {astro-ph.SR},
       adsurl = {https://ui.adsabs.harvard.edu/abs/2023MNRAS.522.2719L}
}

@ARTICLE{Schwope1997,
       author = {{Schwope}, A.~D. and {Mantel}, K.-H. and {Horne}, K.},
        title = "{Phase-resolved high-resolution spectrophotometry of the eclipsing polar HU Aquarii.}",
      journal = {\aap},
         year = 1997,
        month = mar,
       volume = {319},
        pages = {894-908},
          doi = {10.48550/arXiv.astro-ph/9701094},
archivePrefix = {arXiv},
       eprint = {astro-ph/9701094},
 primaryClass = {astro-ph},
       adsurl = {https://ui.adsabs.harvard.edu/abs/1997A&A...319..894S}
}

@ARTICLE{Kuijpers1982,
       author = {{Kuijpers}, J. and {Pringle}, J.~E.},
        title = "{Comments on radial white dwarf accretion}",
      journal = {\aap},
         year = 1982,
        month = oct,
       volume = {114},
       number = {1},
        pages = {L4-L6},
       adsurl = {https://ui.adsabs.harvard.edu/abs/1982A&A...114L...4K}
}

@ARTICLE{Khangale2025,
       author = {{Khangale}, Z.~N. and {Potter}, S.~B. and {Buckley}, D.~A.~H. and {Barrett}, P.~E.},
        title = "{Phase-resolved spectroscopic observations of the magnetic cataclysmic binary EF Eridani: revealing complex magnetic accretion during a high state}",
      journal = {\mnras},
         year = 2025,
        month = nov,
       volume = {544},
       number = {1},
        pages = {309-320},
          doi = {10.1093/mnras/staf1754},
archivePrefix = {arXiv},
       eprint = {2510.08266},
 primaryClass = {astro-ph.SR},
       adsurl = {https://ui.adsabs.harvard.edu/abs/2025MNRAS.544..309K}
}

@INPROCEEDINGS{Beuermann1994,
       author = {{Beuermann}, K. and {Schwope}, A.~D.},
        title = "{AM Herculis Binaries}",
    booktitle = {Interacting Binary Stars},
         year = 1994,
       editor = {{Shafter}, A.~W.},
       series = {Astronomical Society of the Pacific Conference Series},
       volume = {56},
        month = jan,
        pages = {119},
       adsurl = {https://ui.adsabs.harvard.edu/abs/1994ASPC...56..119B}
}

@ARTICLE{Kolbin2020,
       author = {{Kolbin}, A.~I. and {Borisov}, N.~V.},
        title = "{Mapping of White Dwarfs in AM Her Systems}",
      journal = {Astronomy Letters},
         year = 2020,
        month = dec,
       volume = {46},
       number = {12},
        pages = {812-825},
          doi = {10.1134/S1063773720120026},
       adsurl = {https://ui.adsabs.harvard.edu/abs/2020AstL...46..812K}
}

\end{document}